\documentclass[11pt,a4paper]{article}
\usepackage{amsmath,amssymb,amsfonts,a4wide,graphicx,bm,times}
\usepackage{mcite}

\makeatletter \let\old@startsection=\@startsection
\renewcommand{\@startsection}[6]
{\old@startsection{#1}{#2}{#3}{#4}{#5}{#6\mathversion{bold}}}
\makeatother

\makeatletter

\@addtoreset{equation}{section}
\makeatother
\let\refOld\ref
\renewcommand{\ref}[1]{(\refOld{#1})}

\newcommand{\superp}[2]{\genfrac{}{}{0pt}{}{#1}{#2}}

 \def\d{\delta}

 \def\p{\partial}
 
 \def\a{\alpha}
 
 \def\g{\gamma}
 \def\d{\delta}
 \def\e{\varepsilon}

 \def\l{\lambda}

 \def\s{\sigma}

 \def\G{\Gamma}
 \def\D{\Delta}

 \def\L{\Lambda}

\def\CC{{\mathcal{C}}}

\def\CF{{\mathcal{F}}}

\def\CS{{\mathcal{S}}}

\def\CZ{{\mathcal{Z}}}

\def\hf{\dfrac{1}{2}}

\def\implies{\quad\Rightarrow\quad}
\def\equivalent{\quad\Longleftrightarrow\quad}

\def\vphi{\varphi}

\def\CF{\mathcal{F}}

\def\YGz{Y_G^{(0)}}

\def\Ib{\bar I}
\def\hx{\hat x}

\begin{document}
\begin{titlepage}
\renewcommand{\thefootnote}{\fnsymbol{footnote}}
\vspace*{-2cm}
\begin{flushright}
 \jobname .pdf\\ \today
\end{flushright}

\vspace*{1cm}
    \begin{Large}
       \begin{center}
         {\huge Mayer expansion of the Nekrasov prepotential:\\ 
         the subleading $\e_2$-order}
       \end{center}
    \end{Large}
\vspace{0.7cm}

\begin{center}
Jean-Emile Bourgine, Davide Fioravanti\footnote{e-mail address : bourgine@bo.infn.it, fioravanti@bo.infn.it}\\

\vspace{0.7cm}   
{\em Sezione INFN di Bologna, Dipartimento di Fisica e Astronomia,\\
Universit\`a di Bologna} \\
{\em Via Irnerio 46, Bologna, Italy}\\

\end{center}

\vspace{0.7cm}

\begin{abstract}
\noindent
The Mayer cluster expansion technique is applied to the Nekrasov instanton partition function of $\mathcal{N}=2$ $SU(N_c)$ super Yang-Mills. The subleading small $\e_2$-correction to the Nekrasov-Shatashvili limiting value of the prepotential is determined by a detailed analysis of all the one-loop diagrams. Indeed, several types of contributions can be distinguished according to their origin: long range interaction or potential expansion, clusters self-energy, internal structure, one-loop cyclic diagrams, etc.. The field theory result derived more efficiently in \cite{Bourgine2015}, under some minor technical assumptions, receives here definite confirmation thanks to several remarkable cancellations: in this way, we may infer the validity of these assumptions for further computations in the field theoretical approach.
\vspace{0.5cm}
\end{abstract}

\vfill

\end{titlepage}
\vfil\eject

\setcounter{footnote}{0}

%

\section{Introduction}
The Mayer cluster expansion has originally been designed for the computation of the free energy for various fluids under certain physical approximations \cite{Mayer1940,Mayer1941,Andersen1977}. But, as occurred already in other areas of theoretical physics, this methodology turned out recently to be applicable in a very far research topic, namely the Nekrasov partition function. The latter provides exactly all the instanton contributions of $\mathcal{N}=2$ SYM theories with an infinite volume regularisation of $\mathbb{R}^4$, the Omega background, in the computation by localisation \cite{Nekrasov2003}. Eventually, it acquires a simpler multi-integral form which shares some similarities with the grand-canonical partition function of a fluid\footnote{As in many other cases of theoretical physics, the mathematical analogies of the formul{\ae} make the magic.} \cite{Nekrasov2003}. It depends on the two equivariant deformation parameters of the Omega background, $\e_1$ and $\e_2$, and still shows many integrability features of the original Seiberg-Witten theory \cite{SW} (for the integrable hierarchy curve {\it cf.} \cite{Marshakov1999} and references therein). For instance, despite the breaking of Lorentz invariance, nevertheless it exhibits covariance under the Spherical Hecke central (SHc) algebra \cite{Schiffmann2012,Kanno2013} (which is formally equivalent to a $W_\infty$ algebra). The presence of this algebra shed light on the so-called AGT conjecture concerning a duality between these four dimensional theories and the family of Toda conformal field theories in two dimensions \cite{Alday2009,Wyllard2009}\footnote{The conjecture has been partially proved either using the basis of AFLT states \cite{Alba2010}, or a set of generalised Jack polynomials \cite{Morozov2013,Mironov2013}.}. Moreover, the SHc algebra is closely related to a tensorial version of the integrable Calogero-Moser Hamiltonian and led to the construction of one of the most basic objects of quantum integrable theories, namely a (instanton) $R$-matrix \cite{Maulik2012}.

Significantly, in the Nekrasov-Shatashvili (NS) limit $\e_2\to0$, the instanton partition function allows for an effective Mayer expansion which can be summed exactly to the exponential of a Yang-Yang functional: the latter originates, upon minimisation, a suggestive Thermodynamic Bethe Ansatz-like (TBA-like) non-linear integral equation (NLIE) \cite{Nekrasov2009}. All the details of these computations were carried out in \cite{Meneghelli2013,Bourgine2014} where the minimal Yang-Yang functional, the prepotential, has been obtained at leading order along with the non-linear integral equation (NLIE) for the 'classical' motion: if on one side this substitutes the Bethe Ansatz equation within the Yang-Yang approach \cite{nlie}, on the other it is reminiscent of some TBA \cite{Zam-1}. In any case, this is a big signal of integrability which take place here aside to other issues like a quantisation of a Hitchin system associated to the Seiberg-Witten curve, namely a 'quantum' curve \cite{Mironov2010a}. Actually, this curve is, actually, equivalent to a Baxter TQ-relation \cite{Baxter} upon a change of quantum variables \cite{Bourgine2012a}, while the new integrable system is spectral dual to the previous one \cite{Zenkevich2011}. Of course, the TQ-relation is a key object in the integrable model theory and may also be obtained directly by extremising the sum over Young diagrams of the Nekrasov partition function \cite{Poghossian2010}.

Furthermore, in the NS limit, the prepotential undergoes a remarkable physical phenomenon: the formation of bound states of instantons, dubbed 'hadrons' in \cite{Bourgine2014}. Besides, these bound states are described by the meta-clusters introduced below in the Mayer expansion. The technical reason behind this phenomenon is the presence of poles in the integration kernel that hit the integration contour as $\e_2\to0$. As a consequence, a relevant short-range interaction emerges at distances $\sim \e_2$ so to bind tightly many instantons into a bound state. In order to better understand the emergence of bound states in this peculiar limit, we present here the first correction to the prepotential. The method employed here is an alternative to the field theory argument used in \cite{Bourgine2015}. Since the present method relies directly on a double Mayer expansion, with long- and short-range interaction, it explains with more evidence the formation of bound states when $\e_2\to 0$. Moreover, when we depart from this limit, the origin of the different contributions becomes fully detailed, and a systematics at all orders may be put forward. As a technical bonus, this derivation also provides a confirmation of some minor assumptions of the precedent derivation, thus giving it a stronger justification {\it ex juvantibus} \footnote{In the sense that it works well.}, besides the physical ones of \cite{Bourgine2015}. In any case, we would like to suppose that a full disentanglement of the $\e_2$-corrections to the prepotential may pave the way to a better understanding of the full $\e_2$-deformation and its meaning as quantum integrable system characterised by some TBA/NLIE. This should ultimately lead to a richer algebraic structure similar to the SHc Hopf algebra as we know how, on the other hand, the NS limit of the SHc algebra can be obtained \cite{Bourgine2014a}. For simplicity's sake, we shall focus on $\mathcal{N}=2$ SYM with a single $SU(N_c)$ gauge group and a number of fundamental flavours, but the mathematical construction appears to be easily generalisable to arbitrary quivers (possibly along the lines of \cite{Bourgine2014a}). Besides, it should be applicable to 6D theories at least in the specific Omega background $\mathbb{R}^2_{\e_1}\times  \mathbb{R}^2_{\e_2}\times  \mathbb{R}^2_{\e_3}$ when $\e_3\to 0$ (for instance the partition function with some recent developments may be found in \cite{Szabo}).

Eventually, we would like to point out a possible liaison of the present partition function with MHV gluon amplitudes/Wilson loops (WLs) of ${\cal N} = 4$ SYM \cite{AM-amp} as a series \cite{Anope-BSV} because of the mathematical details of the latter. And more evidently a connexion of the NS weak $\Omega$-background $\epsilon_2 \sim 0$ with the string strong coupling  regime $\lambda \gg 1$. These considerations have been put forward in \cite{FPR2, BFPR} and may facilitate the string one-loop computation of the amplitude/WL as $\e_2\sim 1/\sqrt{\l}$; to them we will come back in the last Section {\it Conclusions in perspective}.

This is the paper plan. The general structure of the clusters appearing in the Mayer expansion will be analysed in Section 2. At leading order, only $G$-trees clusters contribute, they also provide an $\e_2$-correction which is derived in the section three. At next-to-leading order, additional clusters contribute, called here $G1$-cycles. The corresponding correction is computed in Section 4. Both types of corrections are combined in Section 5 in order to produce our main result. The specific r\^ole of each term is then briefly discussed. The appendix \refOld{AppA} contains the evaluation of certain type of nested integrals that appear in this problem. In the appendix \refOld{AppC}, clusters of a specific class, the $p$-necklaces, are computed up to the order $O(\e_2)$, as this is necessary to the derivation.

\section{Nekrasov instanton partition function in the NS limit.}
The integral expression derived by N. Nekrasov in \cite{Nekrasov2003} for the $\epsilon$-deformed instanton partition function of $\mathcal{N}=2$ $SU(N_c)$ presents some similarities with the grand canonical partition function of a gas of particles in an external potential $\log Q(x)$ and with an interaction potential $\log K(x)$:
\begin{equation}\label{PF}
\CZ=\sum_{N=0}^\infty \dfrac{\L^N}{N!}\left(\dfrac{\e_+}{\e_1\e_2}\right)^N\int{\prod_{i=1}^NQ(\phi_i)\dfrac{d\phi_i}{2i\pi}\prod_{\superp{i,j=1}{i<j}}^NK(\phi_i-\phi_j)}.
\end{equation}
The main difference lies in the integration contour that is now closed, surrounding the upper half plane and including the real axis but excluding a possible singularity at infinity. The potential $Q(x)$ is assumed to be a rational function, which is the case relevant to Nekrasov partition functions in four dimensions. The r\^ole of the chemical potential is played by $\log \L$, where $\L$ is the exponential of the gauge coupling in conformal theories (zero beta-function, {\it e.g.} $N_f=2N_c$) while a suitable renormalisation of it in asymptotically free theories ($N_f<2N_c$).\footnote{The variables $N_f$ and $N_c$ denote respectively the number of flavours (fundamental massive hypermultiplets) and colours (adjoint gauge multiplet).} The interaction is provided by the kernel 
\begin{equation}
K(x)=\dfrac{x^2(x^2-\e_+^2)}{(x^2-\e_1^2)(x^2-\e_2^2)},
\end{equation}
where the two equivariant deformation parameters $\e_1$ and $\e_2$ are supposed to have a positive imaginary part. Within the nested integrations, the clear fact for the poles at $\phi_i=\phi_j\pm\e_1$ and $\phi_i=\phi_j\pm\e_2$ is to assume $\phi_j$ real. In order to perform the expansion about $\e_2=0$, the key property is the kernel decomposition
\begin{equation}
K(x)=1+\e_2p(x)+\e_2G(x),
\label{Kdecomp}
\end{equation}
with 
\begin{align}
\begin{split}
&p(x)=\a p_0(x),\quad p_0(x)=\dfrac{\e_2}{x^2-\e_2^2},\quad \a=\dfrac{\e_1(2\e_2+\e_1)}{\e_1^2-\e_2^2}=1+2\dfrac{\e_2}{\e_1}+O(\e_2^2),\\
&G(x)=G_0(x)+\e_2G_1(x)+O(\e_2^2),\quad G_0(x)=\dfrac{-2\e_1}{x^2-\e_1^2},\quad G_1(x)=\dfrac1{2\e_1}G_0(x).
\end{split}
\end{align}
In this decomposition, we singled out the poles at $x=\pm\e_2$ responsible for the pinching of the integration contour in the small $\e_2$ limit. The kernel $p(x)$ is interpreted as a very strong 
\textbf{short-range interaction} between instantons.\footnote{The reader should have noticed that in \cite{Bourgine2015} $p_0(x)$ is denoted simply by $p(x)$ and the decomposition of the kernel is also different (factorisation).} It is responsible for the formation of bound states named \textbf{hadrons} in \cite{Bourgine2014}. On the other hand, the $G$ kernel is responsible for a \textbf{long-range interaction}, and at leading order (lo) it can be approximated as an effective interaction between hadrons. All the results presented in this paper readily generalise to arbitrary $\e_2$-independent functions $G_0(x)$, $G_1(x)$ as long as they do not possess any singularity on the integration contour (with particular attention to $x=0$).

In the Seiberg-Witten limit, the logarithm of the partition function is proportional to the regularised volume of $\mathbb{R}^4$, i.e. $\sim 1/\e_1\e_2$. To work with a finite quantity, we define the NS free energy as $\CF=\e_2\log\CZ$. It is given by a double Mayer expansion \cite{Mayer1940,Mayer1941,Andersen1977} of the partition function with two types of links associated to the two functions $p$ and $G$ in the decomposition of the kernel $K$. A generic connected cluster with $l$ vertices will be denoted $C_l$, it is characterized by a set of vertices $V(C_l)$ and two sets of links $E_p(C_l)$ and $E_G(C_l)$ associated to the two components $p$ and $G$ of the kernel $K$. Two vertices of a cluster $C_l$ are connected with at most one link, either a $p$- or a $G$-link. Vertices $i\in V(C_l)$ of the clusters are in correspondence with the integration measures $Q(\phi_i)d\phi_i/2i\pi$, and links $<ij>\in E_p(C_l)$ (resp. $<ij>\in E_G(C_l)$) with the interaction $\e_2p(\phi_{ij})$ (resp. $\e_2G(\phi_{ij})$) where we use the shortcut notation $\phi_{ij}=\phi_i-\phi_j$,
\begin{equation}
\CF=\sum_{l=0}^\infty{q^l\sum_{C_l}\dfrac{\e_2^{-(l-1)}}{\s(C_l)}\int{\prod_{i\in V(C_l)}Q(\phi_i)\dfrac{d\phi_i}{2i\pi}\prod_{<ij>\in E_p(C_l)}\e_2p(\phi_{ij})\prod_{<ij>\in E_G(C_l)}\e_2 G(\phi_{ij})}}.
\end{equation}
In this expression, the gauge coupling has been redefined into $q=\L\e_+/\e_1$\footnote{Note again that a different notation has been used in \cite{Bourgine2015}: $q_\text{there}=e^{-\frac12 k(0)\e_2}q_\text{here}$.}. The symmetry factor $\s(C_l)$ equals the cardinality of the automorphism group of the cluster.

\subsection{General structure of minimal and sub-minimal clusters}

\begin{figure}[!t]
\centering
\includegraphics[width=9cm]{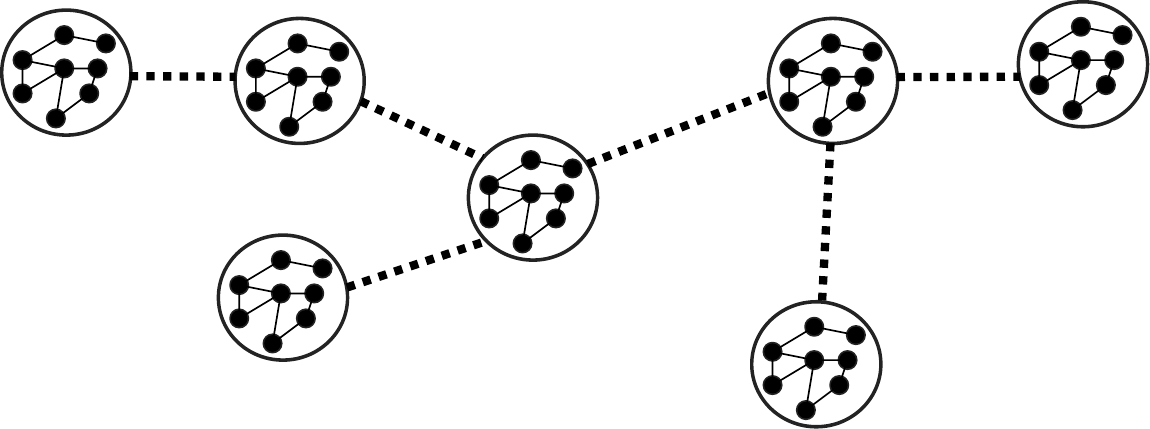}
\caption{General structure of minimal clusters. G-links are dashed and p-links in plain.}
\label{structure}
\end{figure}

Here we repeat and extend the power counting argument employed in \cite{Bourgine2014} to analyze the $\e_2$-order of the clusters integral contributions. To each vertex is attached a factor $\e_2^{-1}$, and $\e_2$ for each $G$-links. Thus, clusters that do not contain $p$-links are \textbf{minimal}, i.e. contribute to the leading order, only if they have a tree structure for which the number of links is minimal and equal the number of vertices minus one. The behavior of $p$-links is more peculiar: when considered in a tree structure they are also of order $O(\e_2)$, but they become of order $O(1)$ if the vertices they link are already connected by a path of $p$-links, thus forming a $p$-cycle. We deduce that minimal clusters are \textbf{$G$-trees}, i.e. clusters without cycles involving G-links. They are represented in the figure \refOld{structure}: a tree structure of $G$-links connects a set of \textbf{meta-vertices} made of sub-clusters of $p$-links. These meta-vertices have been identified with bound-states of instantons, also called hadrons in \cite{Bourgine2014,Bourgine2014a}. By definition, minimal clusters are the only contributions at leading order in $\e_2$, they will be denoted $T_l$. However, they are also responsible for a subleading correction to the prepotential that we denote $\CF_A^{(1)}$ with $\CF^{(1)}=\CF_A^{(1)}+\CF_B^{(1)}$. In order to compute this correction, we use the $G$-trees dressed vertex defined as the generating function of rooted minimal clusters $T_l^x$ with $l$ vertices and a root $x$,
\begin{equation}
Y(x)=Q(x)\sum_{l=1}^\infty{q^l\sum_{T_l^x}\dfrac{\e_2^{-(l-1)}}{\s(T_l^x)}\int{\prod_{\superp{i\in V(T_l^x)}{i\neq x}}Q(\phi_i)\dfrac{d\phi_i}{2i\pi}\prod_{<ij>\in E_p(T_l^x)}\e_2p(\phi_{ij})\prod_{<ij>\in E_G(T_l^x)}\e_2 G(\phi_{ij})}}.
\end{equation}
In this expression, the variable $\phi_x=x$ associated to the root is kept fixed. Due to the presence of this fixed point on the cluster, the group of automorphism preserving the rooted cluster is a subgroup of the group of automorphism acting on the non-rooted one, and the symmetry factors obey $\s(T_l^x)\leq \s(T_l)$. We also introduce the generating functions $Y_G(x)$ and $qQ(x)Y_p(x)$ restricting the summation over clusters $T_l^x$ by requiring the root $x$ to be connected to other vertices only through $G$- or $p$-links respectively. The free energy and the dressed vertex can be $\e_2$-expanded as follows,
\begin{equation}
\CF=\CF^{(0)}+\e_2\CF^{(1)}+O(\e_2^2),\quad Y(x)=Y_0(x)+\e_2Y_1(x)+O(\e_2^2),
\end{equation}
and so on for $Y_G(x)$, $Y_p(x)$,...

\begin{figure}[!t]
\centering
\includegraphics[width=9cm]{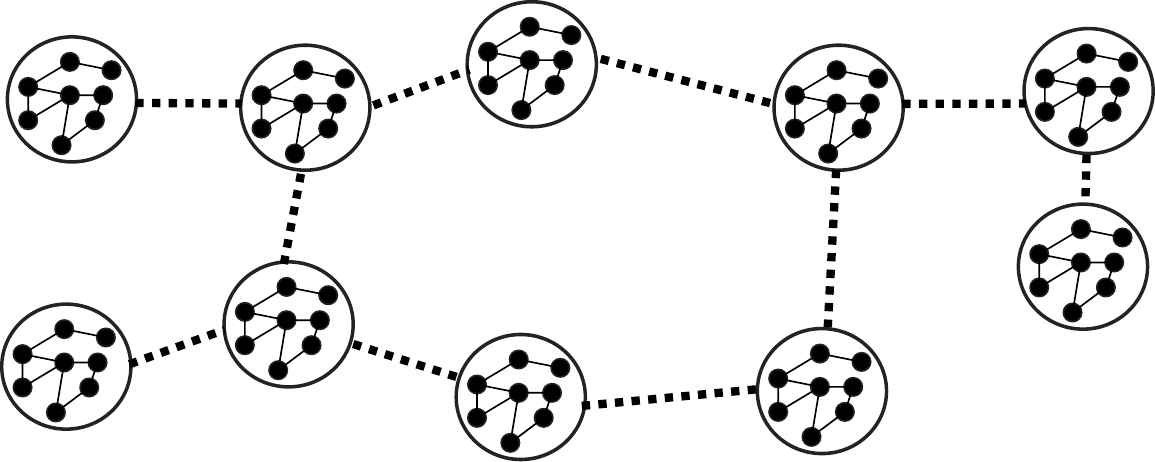}
\caption{General structure of sub-minimal clusters. G-links (dashed) form a necklace involving six dressed meta-vertices.}
\label{metanecklace}
\end{figure}

There are two possibilities to construct the \textbf{sub-minimal} clusters, i.e. the clusters that start contributing at the subleading order in $\e_2$. The first possibility is to add a $G$-link between any two vertices in order to form a cycle involving a $G$-link. Such a cluster will be called a \textbf{$G1$-cycle} since removing a single $G$-link reduces the cluster to a $G$-tree. The second possibility is to add a $p$-link between two vertices that are not connected by a path of $p$-links i.e. belonging to different meta-vertices. In this way, a cycle involving a $G$-link is also formed. This is again a $G1$-cycle since removing a $G$-link produces a minimal cluster. We conclude that the sub-minimal clusters contain a single $G1$-cycle, they have the form of a necklace of meta-vertices dressed by $G$-trees and $p$-clusters (see figure \refOld{metanecklace}). They produce the contribution $\CF_B^{(1)}$ to the free energy.

\subsection{Brief reminder of the results obtained at leading order}
A functional equation characterising the leading order $Y_0(x)$ of the dressed vertex has been established in \cite{Bourgine2014} by exploiting the $G$-tree structure of minimal clusters. This structure implies the factorisation property $Y(x)=Y_p(x)Y_G(x)$ of the generating functions. It is due to the fact that the descendants of vertices related to the root by a $G$-link form independent sub-clusters. The same fact is further responsible for the property
\begin{equation}\label{YG_Y}
Y_G(x)=qQ(x)\exp\left(\int{G(x-y)Y(y)\dfrac{dy}{2i\pi}}\right).
\end{equation}
Considering $Y_p(x)$, the $G$-tree structure also implies the independence of the clusters with roots attached to $x$ via a path of $p$-links, which provides a third relation,
\begin{equation}\label{Yp_YG}
Y_p(x)=\sum_{l=1}^\infty\sum_{\D_l^x}\dfrac{\e_2^{-(l-1)}}{\s(\D_l^x)}\int{\prod_{i\in V(\D_l^x)\smallsetminus\{x\}}Y_G(\phi_i)\dfrac{d\phi_i}{2i\pi}\prod_{<ij>\in E(\D_l^x)}\e_2p(\phi_{ij})},
\end{equation}
where $\D_l^x$ denotes the rooted $p$-cluster that consists of the vertices connected to the root $x$ with a path of $p$-links, and the $p$-links between them that form the set $E(\D_l^x)$. In short, it describes the inner structure of a meta-vertex where $p$-links renders the strong interaction of the instantons that constitute the bound state. It has been shown that at first order the mean field approximation $Y_G(\phi_i)\simeq Y_G(x)$ holds. The remaining series of integrals
\begin{equation}\label{def_Il}
I_l(\a)=\e_2^{-(l-1)}\sum_{\D_l^x}\dfrac1{\s(\D_l^x)}\int{\prod_{i=1}^{l-1}\dfrac{d\phi_i}{2i\pi}\prod_{<ij>\in E(\D_l^x)}\e_2p(\phi_{ij})}
\end{equation}
gives the self-energy of the hadrons, i.e. the energy coming from the $p$-interactions of the instantons in the hadron. Due to the invariance under translation of the variable $\phi_i$, this quantity is actually independent of the mean position $x$ of the constituents. It has been evaluated  at $\a=1$, $I_l(1)=1/l$, which provides the first order in $\e_2$ as $I_l(\a)=I_l(1)+O(\e_2)$. Using the mean field approximation, and the value $1/l$ for the self-energy, we find after multiplication of formula \ref{Yp_YG} by $\YGz(x)$ the following relation:
\begin{equation}\label{rel_Y0_YG0}
Y_0(x)=-\log(1-\YGz(x))\equivalent \YGz(x)=1-e^{-Y_0(x)}.
\end{equation}
Combining it with the relation \ref{YG_Y}, we obtain the celebrated TBA-like \cite{Zam-1} nlie of Nekrasov and Shatashvili \cite{Nekrasov2009}:
\begin{align}
\begin{split}\label{TBA_Y}
&\log(1-e^{-Y_0(x)})=V_0(x)+\int G_0(x-y)Y_0(y)\dfrac{dy}{2i\pi},\\
\text{or}\quad&\YGz(x)=e^{V_0(x)}\exp\left(-\int{G_0(x-y)\log\left(1-\YGz(y)\right)\dfrac{dy}{2i\pi}}\right),
\end{split}
\end{align}
with the expansion of the potential $\log(q Q(x))=V_0(x)+\e_2V_1(x)+O(\e_2^2)$.

Actually, the $G$-tree prepotential may be interpreted as related to the partition function \ref{PF} with special kernel $K(x)$ without the short-range kernel $p(x)$, {\it i.e.} without singularities at $x=\pm\e_2$ to hit the integration contour. It has been already considered in \cite{Bourgine2013} by purely combinatorial means and yields in the present context (with $p(x)$) 
\begin{align}
\begin{split}\label{BSV}
\CF_{G\text{-trees}}=\G_0-\hf\G_1,\quad &\\
&\G_0=\sum_{l=1}^\infty\sum_{\D_l}\dfrac{\e_2^{-(l-1)}}{\s(\D_l)}\int{\prod_{i\in V(\D_l)} Y_G(\phi_i)\dfrac{d\phi_i}{2i\pi}\prod_{<ij>\in E(\D_l)}\e_2p(\phi_{ij})},\\
&\G_1=\int{Y(x)Y(y)G(x-y)\dfrac{dxdy}{(2i\pi)^2}}.
\end{split}
\end{align}
In fact, this formula relates the dressed vertex to the free energy by correcting the symmetry factors associated to the clusters. The coefficient $\G_0$ involves a sum over connected $p$-clusters of $l$ vertices $\D_l$. It is convenient to re-write it as a sum over rooted clusters $\D_l^x$ by exploiting the combinatorial property (B.3) of \cite{Bourgine2013} (without short-range potential):
\begin{equation}\label{G0}
\G_0=\int{\dfrac{dx}{2i\pi}Y_G(x)\sum_{l=1}^\infty\dfrac1l\sum_{\D_l^x}\dfrac{\e_2^{-(l-1)}}{\s(\D_l^x)}\int{\prod_{i\in V(\D_l^x)\smallsetminus\{x\}} Y_G(\phi_i)\dfrac{d\phi_i}{2i\pi}\prod_{<ij>\in E(\D_l^x)}\e_2p(\phi_{ij})}}.
\end{equation}
At first order the mean field approximation $Y_G(\phi_i)\simeq Y_G(x)$ may be used and the resulting integration can be expressed in terms of the self-energy $I_l(\a)$ times an extra symmetry factor $1/l$ due to the indistinguishability of the constituents in the hadron. The calculation of the leading order can be performed exactly, producing
\begin{equation}
\G_0^{(0)}=\int{\dfrac{dx}{2i\pi}\text{Li}_2(1-e^{-Y_0(x)})},\quad \G_1^{(0)}=\int{Y_0(x)Y_0(y)G_0(x-y)\dfrac{dxdy}{(2i\pi)^2}}.
\end{equation}
Now, we can think of this prepotential as the critical value of a an on-shell effective action
\begin{equation}\label{CF0}
\CF^{(0)}=\hf\int{Y_0(x)Y_0(y)G_0(x-y)\dfrac{dxdy}{(2i\pi)^2}}+\int{\dfrac{dx}{2i\pi}\text{Li}_2(\YGz(x))}+\int{\dfrac{dx}{2i\pi}Y_0(x)\left[V_0(x)-\log\YGz(x)\right]}.
\end{equation}
reproducing the nlie \ref{TBA_Y} upon minimisation. In this sense it is dubbed Yang-Yang functional. In fact, the application of the path-integral techniques developed in \cite{FPR2} (for amplitudes) to the present case would simplify the procedure by first summing up all the leading contributions to the Yang-Yang functional \ref{CF0} and then derive \ref{TBA_Y} as associated equation of motion. In specific, this functional approach gives immediately the Yang-Yang functional as in  \cite{Bourgine2013} (first line of \ref{BSV}) without bound states or short-range potential $p(x)$: this is interestingly the TBA for Maxwell-Boltzmann statistics. On the contrary, the addition of bound states, {\it i.e.} short-range potential $p(x)$, modifies the Yang-Yang functional into \ref{CF0} and consequently the stationary equation into the usual TBA \ref{TBA_Y} \cite{Zam-1} with quantum statistics\footnote{As well-known, Bose statistics form can be mapped to the Fermi one.}  \cite{FPR2}.

\section{$G$-trees corrections}
In order to derive the $\e_2$-corrections brought by minimal clusters, we will follow the strategy employed in \cite{Bourgine2014}. First we shall consider the rooted version of the sum over minimal cluster that defines the dressed vertex $Y(x)$. Using various approximation, a linear integral equation, of Fredholm type, will be obtained for the next-to-leading order function $Y_1(x)$. Similar approximations will consequently be made on the formula \ref{BSV} that allows the computation of the prepotential. Remarkably, unlike in the leading order case, it turns out that the terms in the formula \ref{BSV} that involves the dressed potential at subleading order $Y_1(x)$ cancel each-other. As a result, the computation of the prepotential at this order does not require to solve the integral equation of $Y_1(x)$, but can be expressed solely using the leading order quantity $Y_0(x)$.

\subsection{Integral equations}
The restriction of the summation over clusters to $G$-trees in the definition of the dressed vertices $Y(x)$ and $Y_G(x),\ Y_p(x)$ has allowed to derive the functional relations \ref{YG_Y} and \ref{Yp_YG} among them. The first relation can easily be expanded in $\e_2$ and gives at the second order
\begin{equation}\label{equ_YGY}
\dfrac{Y_G^{(1)}(x)}{\YGz(x)}=V_1(x)+\int G_1(x-y) Y_0(y)\dfrac{dy}{2i\pi}+\int G_0(x-y) Y_1(y)\dfrac{dy}{2i\pi}.
\end{equation}
A second relation between $Y_G^{(1)}$ and $Y_1$ should be provided by the expansion of \ref{Yp_YG}. However this equation still contains nested integrations and requires a proper approximation. At the leading order, we have seen that the potential $Y_G(\phi_i)$ attached to the instantons that constitute a hadron can be approximated by a potential at the center of mass $Y_G(x)$, and the remaining integral giving the self-energy of the bound state, denoted $I_l(\a)$, computed in the limit $\e_2\to0$. At the next to leading order, two types of corrections appears. The first type arises due to the fact that the strength of the short range interaction depends on $\e_2$ through the variable $\a=1+2\e_2/\e_1+O(\e_2^2)$. The corresponding correction to the self-energy is computed in appendix \refOld{AppA}, it is given in terms of generalised harmonic numbers,
\begin{equation}\label{exp_Il}
I_l(\a)=I_l(1)+\dfrac{2}{\e_1}I_l'(1)\e_2 +O(\e_2^2),\quad\text{with}\quad I_l'(1)=\hf \sum_{k=1}^{l-1}\dfrac1{k^2}.
\end{equation}

The second contribution of order $O(\e_2)$ is a correction to the mean field approximation $Y_G(\phi_i)\simeq Y_G(x)$ that reflects the internal structure of hadrons. The first move may be to perform a linear approximation of the potentials $Y_G(\phi_i)$ but, as explained in \cite{Bourgine2015}, it is not working because it overlooks the poles of these functions. To properly compute this correction, it is possible to use an approximate interaction kernel $p(x)\simeq p_0(x)$, and borrow the results of this letter obtained for $Y_\text{short}^{(1)}(x)$, replacing the rational potential $U(x)$ by $Y_G(x)$,\footnote{The results presented in the letter \cite{Bourgine2015} rely on two main assumptions. The first one is the possibility to treat the exponential of the Gaussian field as a rational potential. This assumption is not needed here since the formula is applied to the dressed potential $Y_G(x)$ which is a rational function at any finite order in $q$. The second assumption is essential for the derivation, it claims that the total $\e_2$-correction is the sum of contributions brought by each link. It is then shown that only articulation links, i.e. links whose removal break the cluster into two disconnected pieces, provide a non-vanishing contribution. As a consequence, clusters with no articulation links, such as the necklace diagrams computed in the appendix \refOld{AppC}, do not exhibit any $O(\e_2)$ corrections (but higher orders are still present). This second assumption is still needed here.}
\begin{equation}\label{def_D0}
\hf \D_0(x)\nabla Y_0(x),\quad \text{with}\quad \D_0(x)=e^{Y_0(x)}\YGz(x)=e^{Y_0(x)}-1.
\end{equation}
This expression involves a new operator $\nabla$ defined on meromorphic functions that can be split into two parts, $f(x)=f_\text{reg.}(x)+f_\text{sing.}(x)$ where $f_\text{reg.}(x)$ has no poles within the contour of integration, and $f_\text{sing.}(x)$ is a sum of poles inside this contour\footnote{It is the infinitesimal version of the operator $f(x)^+$ that renders the convolution with the $p$-kernel,
\begin{equation}\label{def_plus}
f(x)^+=f_\text{reg.}(x+\e_2)+f_\text{sing.}(x-\e_2)= f(x)+\e_2\nabla f(x)+O(\e_2^2).
\end{equation}}
\begin{equation}\label{def_nabla}
\nabla f(x)=f_\text{reg.}'(x)-f_\text{sing.}'(x).
\end{equation}

Combining the two types of $\e_2$-corrections, and after multiplying the equation \ref{Yp_YG} with $Y_G(x)$ to form $Y(x)$ in the LHS, we derive
\begin{equation}\label{corr_Yp}
Y(x)=\sum_{l=1}^\infty I_l(\a)Y_G(x)^l+\hf\e_2\D_0(x)\nabla Y_0(x)+O(\e_2^2).
\end{equation}
Expanding the self-energy as in \ref{exp_Il}, and inserting the generating functions
\begin{equation}
l_1(x)=\sum_{l=1}^{\infty}I_l(1)x^l=-\log(1-x),\quad \d l_1(x)=\sum_{l=1}^\infty{I_l'(1)x^l}=\dfrac{x}{2}\dfrac{\text{Li}_2(x)}{1-x},
\end{equation}
we finally find the relation
\begin{equation}\label{equ_YYG}
Y_1(x)= e^{Y_0(x)}Y_G^{(1)}(x)+\dfrac{1}{\e_1}\D_0(x)\text{Li}_2\left(\D_0(x)e^{-Y_0(x)}\right)+\hf \D_0(x)\nabla Y_0(x).
\end{equation}

By combining \ref{equ_YGY} and \ref{equ_YYG}, it is possible to write a single functional equation, either for $Y_1(x)$ or for $Y_G^{(1)}(x)$:
\begin{equation}
\dfrac{Y_1(x)}{\D_0(x)}=F(x)+\int{G_0(x-y)Y_1(y)\dfrac{dy}{2i\pi}},\quad \dfrac{Y_G^{(1)}(x)}{\YGz(x)}=F_G(x)+\int{G_0(x-y)\dfrac{Y_G^{(1)}(y)}{\YGz(y)}\D_0(y)\dfrac{dy}{2i\pi}}.
\end{equation}
with the following functions that can be computed using the expression of $Y_0(x)$ obtained at first order,
\begin{align}
\begin{split}
&F(x)=V_1(x)+\int{G_1(x-y)Y_0(y)}\dfrac{dy}{2i\pi}+\hf\nabla Y_0(x)+\dfrac{1}{\e_1}\text{Li}_2\left(\D_0(x)e^{-Y_0(x)}\right),\\
&F_G(x)=V_1(x)+\int{G_1(x-y)Y_0(y)\dfrac{dy}{2i\pi}}+\int{G_0(x-y)\left[\hf\nabla Y_0(y)+\dfrac{1}{\e_1}\text{Li}_2\left(\D_0(y)e^{-Y_0(y)}\right)\right]\D_0(y)\dfrac{dy}{2i\pi}}.
\end{split}
\end{align}
These equations are of Fredholm type and can be solved by iterations,
\begin{align}
\begin{split}
&\dfrac{Y_1(x)}{\D_0(x)}=F(x)+\sum_{n=1}^\infty\int{G_0(x-z_1)\prod_{i=1}^{n-1}G(z_i-z_{i+1})\ F(z_n)\prod_{i=1}^n\D_0(z_i)\dfrac{dz_i}{2i\pi}},\\
&\dfrac{Y_G^{(1)}(x)}{\YGz(x)}=F_G(x)+\sum_{n=1}^\infty\int{G_0(x-z_1)\prod_{i=1}^{n-1}G(z_i-z_{i+1})\ F_G(z_n)\prod_{i=1}^n\D_0(z_i)\dfrac{dz_i}{2i\pi}}.
\end{split}
\end{align}

\paragraph{Regular dressed potential} As a side remark, let us mention that in the case where $Y_G(x)$ has no pole in the integration contour it is possible to simplify the relation \ref{Yp_YG} using the linear approximation $Y_G(\phi_i)\simeq Y_G(x)+(\phi_i-x)\p_xY_G(x)$, which leads to (after multiplication by $Y_G(x)$)
\begin{equation}
Y(x)= \sum_{l=1}^\infty\left[I_l(\a)Y_G(x)^l+\e_2J_l(\a)Y_G(x)^{l-1}\p_xY_G(x)\right]+O(\e_2^2),
\end{equation}
with a new series of nested integrals
\begin{equation}\label{def_Jl}
J_l(\a)=\e_2^{-l}\sum_{\D_l^x}\dfrac1{\s(\D_l^x)}\int{\prod_{i=1}^{l-1}\dfrac{d\phi_i}{2i\pi}\prod_{<ij>\in E(\D_l^x)}\e_2p(\phi_{ij})\ \ \sum_{k=1}^{l-1}(\phi_k-x)}.
\end{equation}
The quantity $J_l(1)$ is computed in the appendix \refOld{AppA}, $J_l(1)=(l-1)/2$, which gives for $Y_1(x)$ the formula
\begin{equation}
Y_1(x)=Y_G^{(1)}(x) e^{Y_0(x)}+\dfrac{1}{\e_1}\D_0(x)\text{Li}_2\left(1-e^{-Y_0(x)}\right)+\hf \D_0(x) \p_x Y_0(x).
\end{equation}
This is equivalent to \ref{equ_YYG} with the replacement $\nabla Y_0(x)\to \p_x Y_0(x)$.

\subsection{Corrections to the formula \ref{BSV}}
To determine the $G$-trees corrections $\CF_A^{(1)}$ to the prepotential, we now perform the $\e_2$-expansion of the formula \ref{BSV}. The expansion of the term $\G_1$ is trivial,
\begin{equation}
\G_1^{(1)}=\int{Y_0(x)Y_0(y)G_1(x-y)\dfrac{dxdy}{(2i\pi)^2}}+2\int{Y_0(x)Y_1(y)G_0(x-y)\dfrac{dxdy}{(2i\pi)^2}}.
\end{equation}
Using the equations of motion \ref{equ_YGY}, this can also be written as
\begin{equation}
\G_1^{(1)}=-\int{Y_0(x)Y_0(y)G_1(x-y)\dfrac{dxdy}{(2i\pi)^2}}-2\int{V_1(x)Y_0(x)\dfrac{dx}{2i\pi}}+2\int{\dfrac{Y_0(x)}{1-e^{-Y_0(x)}}Y_G^{(1)}(x)\dfrac{dx}{2i\pi}}
\end{equation}

The expression \ref{G0} of $\G_0$ is more involved, but similar to the one giving $Y_p(x)$ in \ref{Yp_YG}, up to the extra factor $1/l$. At the order $O(\e_2)$, it exhibits both self-energy and internal structure corrections. The former are again encoded in $I_l(\a)$ whereas the latter can be extracted from the result of \cite{Bourgine2015} for $\CZ_\text{short}[X]$, with the $G$-dressed vertex $Y_G(x)$ playing the role of the potential $U(x)$. Combining the two types of corrections, we arrive at
\begin{equation}\label{corr_G0}
\G_0= \int{\sum_{l=1}^\infty\dfrac1l\left[I_l(\a)Y_G(x)^{l-1}+\dfrac14\e_2Y_G(x)^{l-1}\nabla Y(x)\right]Y_G(x)\dfrac{dx}{2i\pi}}+O(\e_2^2).
\end{equation}
The first term can be expressed in terms of the (logarithmic) primitives of the functions $l_1(x)$ and $\d l_1(x)$,\footnote{Working at finite $\a$, we may define
\begin{equation}
l_\a(x)=\sum_{l=1}^\infty I_l(\a)x^l,\quad L_\a(x)=\sum_{l=1}^\infty\dfrac{I_l(\a)}{l}x^l,
\end{equation}
such that $x\p_xL_\a(x)=l_\a(x)$. Expanding around $\a=1$, we recover
\begin{equation}
l_{1+\g}(x)=l_1(x)+\g\d l_1(x)+O(\g^2),\quad L_{1+\g}(x)=\text{Li}_2(x)+\g\d L_1(x)+O(\g^2).
\end{equation}}
\begin{equation}\label{def_dL1}
L_1(x)=\text{Li}_2(x),\quad\d L_1(x)=\text{Li}_3(1-x)+\hf\text{Li}_1(x)\left(\text{Li}_2(1-x)+\dfrac{\pi^2}{6}\right),
\end{equation}
it gives in terms of $Y_0(x)$:
\begin{equation}
\G_0^{(1)}=\int{\dfrac{dx}{2i\pi}\left[\dfrac{Y_0(x)}{1-e^{-Y_0(x)}}Y_G^{(1)}(x)+\dfrac{2}{\e_1}\d L_1\left(1-e^{-Y_0(x)}\right)+\dfrac14 Y_0(x)\nabla Y_0(x)\right]}.
\end{equation}
Finally, summing both contributions from $\G_0$ and $\frac12\G_1$, we observe a cancellation among the terms involving the subleading $G$-dressed vertex $Y_G^{(1)}(x)$, and the $G$-trees part of the prepotential reads
\begin{equation}
\CF_A^{(1)}=\int{\dfrac{dx}{2i\pi}\left[V_1(x)Y_0(x)+\dfrac{2}{\e_1}\d L_1(1-e^{-Y_0(x)})+\dfrac14 Y_0(x)\nabla Y_0(x)\right]}+\hf \int{Y_0(x)Y_0(y)G_1(x-y)\dfrac{dxdy}{(2i\pi)^2}}.
\end{equation}
It no longer depends on the $\e_2$-corrected quantities $Y_1(x)$ and $Y_G^{(1)}(x)$. In this expression, each term has a clear interpretation. The first term corresponds to the expansion of the potential $Q(x)$ that involves the $\e_2$ parameter. The second and last terms come from the dependence of the interactions ($p$ and $G$ kernels respectively) in $\e_2$. The remaining term involving the operator $\nabla$ can be traced back to the internal structure corrections. 

\section{$G1$-cycles corrections}
By definition, the sub-minimal clusters start contributing to the prepotential at the order $O(\e_2)$. As a consequence, $\e_2$-corrections to the self-energy and internal structure become subleading and will be neglected. It has been argued that these clusters have the structure presented in the figure \refOld{metanecklace}, and thus contain a $G1$-cycle, i.e. a closed path involving at least one $G$-link. The number of $G$-links in the $G1$-cycle defines a grading of the sub-minimal clusters that we call the \textbf{order}.

Removing a $G$-link from a sub-minimal cluster gives either a $G$-tree (if the link belongs to the $G1$-cycle) or two disconnected pieces (otherwise). Consider the first case, i.e. a link from the $G1$-cycle, and denote $x$ and $y$ its two extremities. The cluster obtained by removing the $G$-link is a bi-rooted $G$-tree denoted $T_l^{x,y}$. Reversely, any sub-minimal cluster can be engineered from a bi-rooted $G$-tree by inserting a $G$-link between the two roots. Then, of course, the symmetry factors must be adjusted properly. It will thus be useful to introduce the generating function of bi-rooted $G$-trees, or $G$-trees dressed propagator,
\begin{equation}
Y(x,y)=Q(x)Q(y)\sum_{l=2}^\infty{q^l\sum_{T_l^{x,y}}\dfrac{\e_2^{-(l-1)}}{\s(T_l^{x,y})}\int{\prod_{\superp{i\in V(T_l^{x,y})}{i\neq x,y}}Q(\phi_i)\dfrac{d\phi_i}{2i\pi}\prod_{<ij>\in E_p(T_l^{x,y})}\e_2p(\phi_{ij})\prod_{<ij>\in E_G(T_l^{x,y})}\e_2 G(\phi_{ij})}}.
\end{equation}
The first summation is over the total number $l$ of vertices in the clusters $T_l^{x,y}$, including the two roots $x$ and $y$, hence $l\geq 2$. It is also possible to define on the bi-rooted $G$-trees $T_l^{x,y}$ a grading similar to the order previously defined for the sub-minimal clusters. Assuming that there exists a path from $x$ to $y$ involving a $G$-link, then the number of $G$-links in this path is fixed for a given cluster (otherwise there would exist a $G$-cycle). This constant is called the order of the bi-rooted tree. By extension, a bi-rooted $G$-tree of order zero links $x$ to $y$ with only $p$-links. Note that in the process of removing a $G$-link from the $G1$-cycle of a sub-minimal cluster, the order decreases by one.

An important fact about the bi-rooted $G$-trees is that their generating function can be obtained from the generating function $Y(x)$ of trees with a single root by a functional derivation,
\begin{equation}\label{diff_Y}
2i\pi Q(y)\dfrac{\d Y(x)}{\d Q(y)}=Y(x,y)+2i\pi\d(x-y) Y(x).
\end{equation}
In particular, the functional equation obeyed by $Y_0(x)$ \ref{TBA_Y} at first order in $\e_2$ induces a functional relation on its bi-rooted version $Y_0(x,y)$. By a careful analysis of this equation we will be able to obtain the generic term for contribution of sub-minimal clusters of order $n$ to the prepotential. To illustrate our approach, we first recall the case $\a=0$, studied in \cite{Bourgine2013} and for which the clusters have no $p$-links.

\subsection{Sub-minimal clusters with only $G$-links}
As written above, the case of the kernel $K(x)$ with no singularities at $x=\pm\e_2$ has been treated in \cite{Bourgine2013}. Here, it corresponds to the case $\a=0$ or $p(x)=0$. In this case, the prepotential has been obtained up to the order $O(\e_2)$ by various methods. We now present the simplest method that will later be generalised to the case $\a\neq 0$.

When there are no $p$-links, the dressed vertex $Y_0(x)$ satisfies a simplified integral equation, sometimes referred as a linearised form of TBA,
\begin{equation}\label{lin_TBA_Y}
\log Y_0(x)=V_0(x)+\int{\dfrac{dy}{2i\pi}G_0(x-y)Y_0(y)},
\end{equation}
with $V_0(x)$ the leading order in the $\e_2$-expansion of the logarithmic potential $\log qQ(x)$. Taking the differential with respect to the potential of an exponentiated version of this equation, and then using the property \ref{diff_Y}, we deduce a functional equation satisfied by the tree-level dressed propagator,
\begin{equation}\label{equ_Y0_a0}
Y_0(x,y)=Y_0(x)Y_0(y)G_0(x-y)+Y_0(x)\int\dfrac{dz}{2i\pi}G_0(x-z)Y_0(z,y).
\end{equation}
This equation is a Fredholm equation of the second type. It can be solved by iterations, producing the Liouville-Neumann series\footnote{This series appears as the inversion of
\begin{equation}
Y_0(x,y)=2i\pi Y_0(x)\left[\dfrac1{\d(x-y)-\frac1{2i\pi}G(x-y)Y_0(y)}-\d(x-y)\right],
\end{equation}
which clarifies the interpretation of the bi-rooted generating function as a propagator.}
\begin{equation}
Y_0(x,y)=Y_0(x)Y_0(y)\left[G_0(x-y)+\sum_{n=1}^\infty\int{\prod_{i=1}^nY_0(z_i)\dfrac{dz_i}{2i\pi}\ G_0(x-z_1)\prod_{i=1}^{n-1}G_0(z_i-z_{i+1})\ G_0(z_n-y)}\right].
\end{equation}
This series has an important graphical interpretation: the $n$th term corresponds to clusters of order $n+1$ in which the roots are connected by a path of $n+1$ $G$-links. This path contains $n$ intermediate vertices $z_1,\cdots ,z_n$ all dressed by $G$-trees due to the presence of the dressed potentials $Y_0(z_i)$.

The subleading correction to the prepotential $\CF^{(1)}=\CF_B^{(1)}$ takes the form of a similar series, but with an extra link between the roots $x$ and $y$, and an integration over these variables. The resulting series is a sum over necklaces, i.e. clusters of $n$ vertices and $n$ links as depicted in figure \refOld{necklace}a below, with their vertices again dressed by $G$-trees. The symmetry factor $2n$ is given by the cardinal of the dihedral group that acts on the necklace,
\begin{equation}\label{CF1_a0}
\CF^{(1)}=\sum_{n=3}^\infty\dfrac1{2n}\int{\prod_{i=1}^nY_0(\phi_i)\dfrac{d\phi_i}{2i\pi}\prod_{i=1}^{n-1}G_0(\phi_i-\phi_{i+1}) G_0(\phi_n-\phi_1)}.
\end{equation}
The sum starts at $n=3$ since at least three vertices are needed to form a cycle. We recognize the expansion of a Fredholm determinant where the first two terms went missing,
\begin{equation}\label{CF1_no_p}
\CF^{(1)}=-\hf\log\det\left[\d(x-y)-\frac1{2i\pi}G_0(x-y)Y_0(y)\right]-\hf G_0(0)\int\dfrac{dx}{2i\pi}Y_0(x)-\dfrac14\int\dfrac{dxdy}{(2i\pi)^2}Y_0(x)Y_0(y)G_0(x-y)^2.
\end{equation}
The two missing terms are interpreted as degenerate configurations excluded from the Mayer expansion. The term of order one is a tadpole, i.e. a link looping on a single (dressed) vertex. The term of order two has two vertices and two links.

\subsection{Bi-rooted $G$-trees with $p$-links}
The short range interaction $p$ modifies the simple NLIE \ref{lin_TBA_Y} into the more involved TBA-like NLIE \ref{TBA_Y}. The equation for the dressed propagator can still be obtained by a functional differentiation and reads
\begin{equation}
Y_0(x,y)=2i\pi\d(x-y)\left(\D_0-Y_0(x)\right)+G_0(x-y)\D_0(x)Y_0(y)+\D_0(x)\int{G_0(x-z)Y_0(z,y)\dfrac{dz}{2i\pi}},
\end{equation}
where we used the function $\D_0(x)$ defined in \ref{def_D0}. The functional relation takes a simpler form if we define
\begin{equation}
\bar Y_0(x,y)=Y_0(x,y)+2i\pi\d(x-y)\left[Y_0(x)-\D_0(x)\right],
\end{equation}
which satisfies
\begin{equation}
\bar Y_0(x,y)=\D_0(x)\D_0(y)G_0(x-y)+\D_0(x)\int{G_0(x-z)\bar Y_0(z,y)\dfrac{dz}{2i\pi}}.
\end{equation}
This equation is the same as \ref{equ_Y0_a0} obtained in the case $\a=0$, with $Y_0(x)$ replaced by $\D_0(x)$. Its solution is still given by the Liouville-Neumann series
\begin{equation}\label{bY0_sol}
\bar Y_0(x,y)=\D_0(x)\D_0(y)\left[G_0(x-y)+\sum_{n=1}^\infty\int{\prod_{i=1}^n\D_0(z_i)\dfrac{dz_i}{2i\pi}G_0(x-z_1)\prod_{i=1}^{n-1}G_0(z_i-z_{i+1})G_0(z_n-y)}\right].
\end{equation}

\begin{figure}[!t]
\centering
\includegraphics[width=13cm]{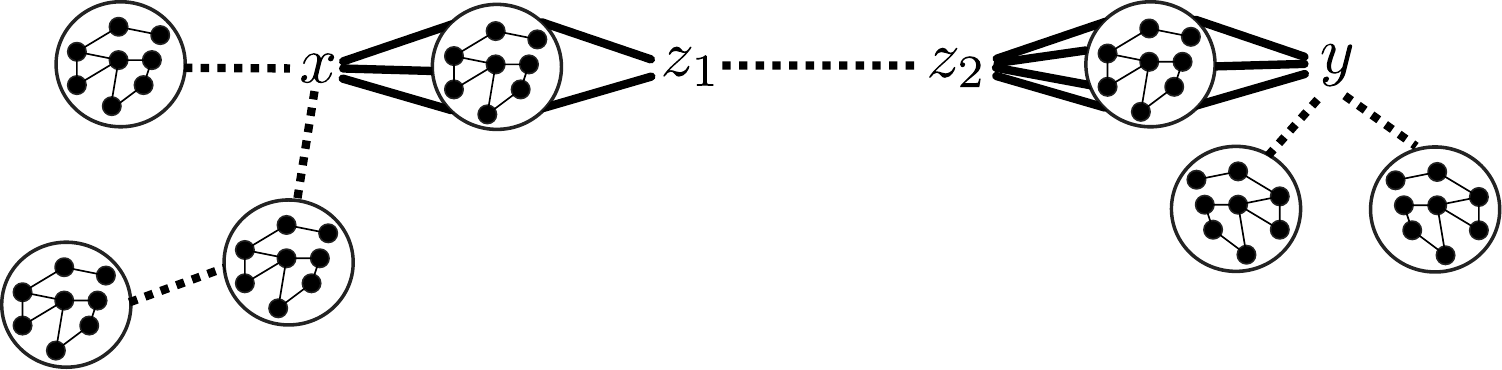}
\caption{Bi-rooted $G$-tree of order one, with roots $x$ and $y$ and extremities $z_1$ and $z_2$ of the $G$-link (dashed).}
\label{biorder1}
\end{figure}

In order to understand the effect of the $p$-links on the functional equation and the resulting replacement of the dressing potential $Y_0(x)$ with $\D_0(x)$, we first consider a bi-rooted cluster of order one, and denote $z_1$ and $z_2$ the extremities of the $G$-link in the path connecting $x$ to $y$ (see figure \refOld{biorder1}). This type of clusters is built from two rooted $p$-clusters $\D^{x}$ and $\D^{y}$ by inserting a $G$-link between two of their vertices, and then dressing all vertices by $G$-trees. It implies that the contribution of clusters of order one to $Y_0(x,y)$ reads, up to $\e_2$-corrections,
\begin{equation}
\sum_{\D^x,\D^y}\dfrac{\sharp V(\D^x)\times \sharp V(\D^y)}{\s(\D^x)\s(\D^y)}\int\prod_{\superp{i\in V(\D^{x})\cup V(\D^{y})}{i\neq x,y}}Y_G(\phi_i)\dfrac{d\phi_i}{2i\pi}\prod_{<ij>\in E_p(\D^{x})\cup E_p(\D^{y})}\e_2p(\phi_{ij})\ \ \ G(z_1-z_2).
\end{equation}
The extra factor $\sharp V(\D^x)\times \sharp V(\D^y)$ corresponds to the choice of a vertex in $\D^x$ and $\D^y$ to form the $G$-link. The remaining symmetry factor is simply the product of the symmetries of the constituents $\D^x$ and $\D^y$: since the roots are fixed by automorphisms, it is not possible to enhance the symmetry when adding the $G$-link between $z_1$ and $z_2$. At first order in $\e_2$, the potentials at each vertex of a $p$-cluster can be replaced by an average potential at one of the roots. It follows that the previous contribution takes the factorised form $\D(x)\D(y)G(x-y)$ with $\D(x)$ a sum over bi-rooted $p$-clusters with $l$ vertices dressed by $Y_G$,
\begin{equation}
\quad \D(x)=\sum_{l=1}^\infty lY_G(x) \sum_{\D_l^{x}}\dfrac1{\s(\D_l^x)}\int\prod_{\superp{i\in V(\D^x)}{i\neq x}}Y_G(\phi_i)\dfrac{d\phi_i}{2i\pi}\prod_{<ij>\in E_p(\D^x)}\e_2p(\phi_{ij}).
\end{equation}
Performing the usual approximation $Y_G(\phi_i)\simeq Y_G(x)$, we recognize $I_l(\a)$ in the remaining sum of integrals, so that at first order in $\e_2$, $\D(x)=\D_0(x)+O(\e_2)$ with
\begin{equation}
\D_0(x)=\sum_{l=1}^\infty \left(Y_G^{(0)}(x)\right)^l=e^{Y_0(x)}-1.
\end{equation}
Of course, this result coincides with the previous definition of $\D_0(x)$ given in \ref{def_D0}. We have thus reconstructed the contribution of order one to $Y_0(x,y)$: $\D_0(x)\D_0(y)G_0(x-y)$. By induction, we now understand the role of the $p$-links that dress the potential $Y_0(x)$ into $\D_0(x)$ by substituting meta-vertices to $G$-trees as dressing factors of vertices in the path joining $x$ to $y$. These meta-vertices are represented by intermediate $p$-clusters with inner vertices dressed by $G$-trees. Consequently, the Neumann-Liouville series \ref{bY0_sol} can still be interpreted as an expansion over the order of the bi-rooted clusters.

It remains to explain the difference between the dressed propagator $Y(x,y)$ and $\bar Y(x,y)$. Since the difference is proportional to the delta function $\d(x-y)$, it only affects the degenerate terms in which the two roots coincide. As in the case $\a=0$ treated in \cite{Bourgine2013}, the original dressed propagator $Y(x,y)$ has a singularity $Y(x,y)\sim -2i\pi Y(x)\d(x-y)$ as $x\to y$. This singularity is dressed by the presence of $p$-links in $\bar Y(x)$ where now $\bar Y(x,y)\sim -2i\pi \D(x)\d(x-y)$.

\subsection{Contribution to the prepotential}
\begin{figure}[!t]
\centering
\includegraphics[width=13cm]{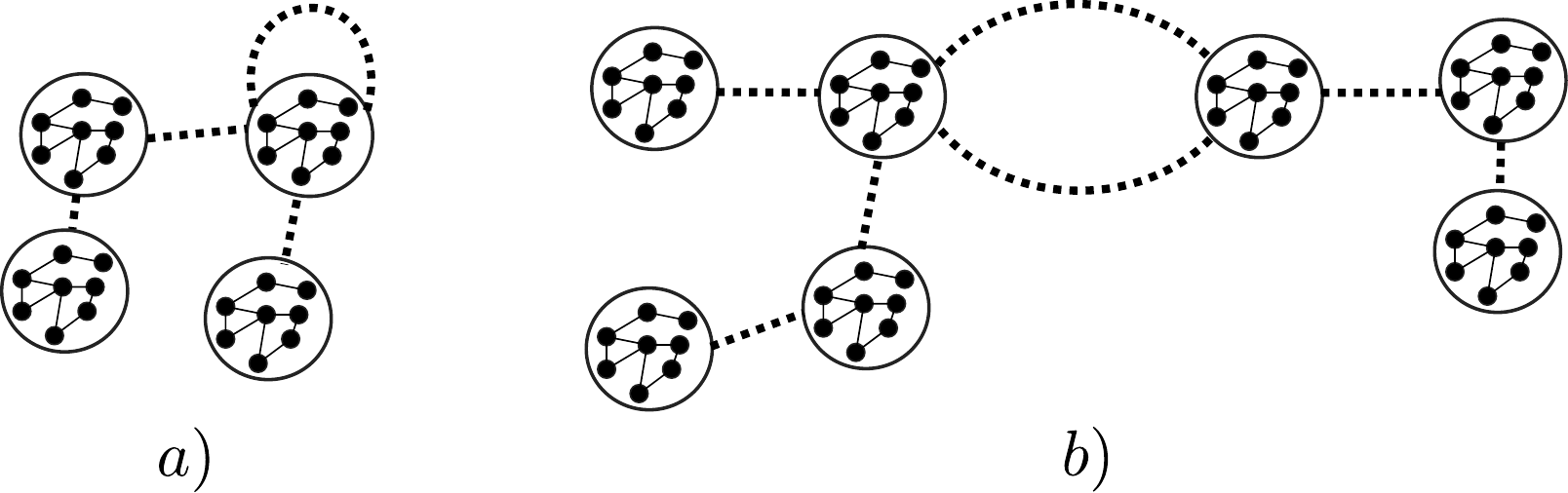}
\caption{a) A sub-minimal cluster of order one. b) A sub-minimal cluster of order two.}
\label{order2}
\end{figure}

It is now safe to postulate that the generic contribution of sub-minimal clusters of order $n$ to the prepotential is given by
\begin{equation}
\dfrac1{2n}\int{\prod_{i=1}^n\D_0(\phi_i)\dfrac{d\phi_i}{2i\pi}\prod_{i=1}^{n-1}G_0(\phi_i-\phi_{i+1})\ G_0(\phi_n-\phi_1)}.
\end{equation}
This result is valid for $n\geq 3$. As for the toy-model $\a=0$, it is necessary to subtract some degenerate configurations at low order. To determine the associated contributions, we come back to the construction of sub-minimal clusters by adding a link to a bi-rooted $G$-tree. At the order two, we add a link between two meta-vertices that are already connected by a $G$-link, thus obtaining a configuration of two meta-vertices connected by two $G$-links (see figure \refOld{order2}). This configuration is now allowed, provided that the extremities of the two $G$-links do not coincide. The contribution of this degenerate configuration is the same as in the case $\a=0$, namely
\begin{equation}
\dfrac14\int\dfrac{dxdy}{(2i\pi)^2}Y_0(x)Y_0(y)G_0(x-y)^2.
\end{equation}

When no p-links are present, the $G$-cycles of order one correspond to a tadpole configuration where a link loops back to its initial vertex (possibly dressed by $G$-trees). These configurations are not allowed in the Mayer expansion. Introducing the $p$-links, vertices of $G$-trees are replaced by meta-vertices corresponding to $p$-clusters (or hadrons). A $G$-cycle of order one now corresponds to a G-link inserted between two vertices of the same meta-vertex, i.e. already related through a path involving only $p$-links. These configurations are now permitted provided that
\begin{enumerate}
 \item The two vertices at the extremities of the G-link are different.
 \item These two vertices are not already linked together with a (single) $p$-link.
\end{enumerate}
Instead of enumerating the configurations that are not permitted, it is simpler to perform a direct calculation of the contributions of $G1$-cycles of order one. Let us denote $x$ and $y$ the two extremities of the $G$-link, and $\bar\D_l^{x,y}$ the bi-rooted $p$-cluster corresponding to the meta-vertex containing $x$ and $y$. In this bi-rooted $p$-cluster, the roots $x$ and $y$ cannot be directly connected with a $p$-link, and this additional constraint is represented with the bar in our notation $\bar\D_l^{x,y}$. Each vertex of the $p$-cluster can be dressed with G-trees, and the contribution to the prepotential reads
\begin{equation}
\hf\int{\dfrac{dxdy}{(2i\pi)^2}G(x-y)Y_G(x)Y_G(y)\sum_{l=3}^\infty\e_2^{2-l}\sum_{\bar\D_l^{x,y}}\dfrac1{\s(\bar\D_l^{x,y})}\int{\prod_{\superp{i\in V(\bar\D_l^{x,y})}{x\neq i\neq y}}Y_G(\phi_i)\dfrac{d\phi_i}{2i\pi}\prod_{<ij>\in E_p(\bar\D_l^{x,y})}\e_2 p(\phi_i-\phi_j)}}.
\end{equation}
The symmetry factor $1/2$ in front corresponds to exchange the role of $x$ and $y$. Note also that at least three vertices are required in the meta-vertex in order to be able to form a $G1$-cycle. Consequently, in the first sum the number $l$ of vertices is always greater or equal to three. At the leading order in $\e_2$, it is possible to do the approximation $Y_G(x)\simeq Y_G(y)\simeq Y_G(\phi_i)$ for the inner vertices of a meta-vertex. The extra $G$-link can be treated as a potential in this approximation, $G(x-y)\simeq G(0)$. The contributions of order one become
\begin{equation}
\hf G(0)\int{\dfrac{dxdy}{(2i\pi)^2}\sum_{l=3}^\infty Y_G(x)^l \Ib_l^{x,y}},
\end{equation}
where the remaining integral is defined and evaluated in the appendix \refOld{AppB}. In particular, \ref{int_Ib} gives the $y$-integral of $\Ib_l^{x,y}$ in terms of generalised harmonic numbers. Re-introducing the function $\d L_1(x)$ defined in \ref{def_dL1} allows us to write the terms of order one as
\begin{equation}
\hf\e_2G(0)\int{\dfrac{dx}{2i\pi}\left[\dfrac1{1-Y_G(x)}+\log(1-Y_G(x))-2\d L_1(Y_G(x))\right]}.
\end{equation}
At first order in $\e_2$, and expressed in terms of $Y_0(x)$ with $G_0(0)=2/\e_1$, it becomes
\begin{equation}
\dfrac{\e_2}{\e_1}\int{\dfrac{dx}{2i\pi}\left[\D_0(x)-Y_0(x)-2\d L_1\left(1-e^{-Y_0(x)}\right)\right]}.
\end{equation}
The first term is what we would expect if all the tadpole configurations were permitted. The second term cancel the configuration where the meta-vertex is actually a single vertex (dressed by $G$-trees). More interestingly, the last term is specific to our problem and cancel part of the self-energy $\e_2$-correction in $\CF_A^{(1)}$.

Combining all order contributions of the $G1$-cycles, we write
\begin{align}
\begin{split}
\CF_B^{(1)}&=\sum_{n=1}^\infty\dfrac1{2n}\int{\prod_{i=1}^n\D_0(\phi_i)\dfrac{d\phi_i}{2i\pi}\prod_{i=1}^{n-1}G_0(\phi_i-\phi_{i+1}) G_0(\phi_n-\phi_1)}\\
&-\hf G_0(0)\int\dfrac{dx}{2i\pi}Y_0(x)-G_0(0)\int{\d L_1\left(1-e^{-Y_0(x)}\right)}-\dfrac14\int\dfrac{dxdy}{(2i\pi)^2}Y_0(x)Y_0(y)G_0(x-y)^2.
\end{split}
\end{align}
The sum in the first line coincides with the expansion of a Fredholm determinant.

\section{Conclusions in perspective}
Re-assembling $G$-trees and $G1$-cycles corrections, we find that the finite $\e_2$-contributions to the prepotential takes the following form,
\begin{align}
\begin{split}\label{final}
\CF^{(1)}&=-\hf\log\det\left[\d(x-y)-\frac1{2i\pi}G_0(x-y)\D_0(y)\right]+\int{\dfrac{dx}{2i\pi}\left[V_1(x)Y_0(x)-\dfrac1{\e_1}Y_0(x)\right]}\\
&+\hf \int{Y_0(x)Y_0(y)G_1(x-y)\dfrac{dxdy}{(2i\pi)^2}}-\dfrac14\int\dfrac{dxdy}{(2i\pi)^2}Y_0(x)Y_0(y)G_0(x-y)^2+\dfrac14\int{\dfrac{dx}{2i\pi}Y_0(x)\nabla Y_0(x)}.
\end{split}
\end{align}
In order to make contact with the result presented in \cite{Bourgine2015}, and based on a field theory description, the terms above have to be re-arranged. First, the two kinetic terms, quadratic in $Y_0(x)$, can be combined to form the $\e_2$-expansion of a new kernel $k(x)$ that appears in the factorisation of the full original kernel $K(x)$ into short- and long-range interactions:
\begin{equation}
K(x)=(1+\e_2p_0(x))e^{\e_2 k(x)},\quad k(x)=G_0(x)+\e_2\left(G_1(x)-\hf G_0(x)^2\right)+O(\e_2^2).
\end{equation}
The latter factorisation\footnote{In contrast to the linear decomposition $K=1+\e_2p+\e_2G$ employed in this paper.} turned out to be better suited for the separation of the two kinds of interactions. In the Mayer expansion formalism, the two terms that compose the $\e_2$-expansion of the kernel $k(x)$ at the order $O(\e_2)$ comes from diagrams containing two meta-vertices, or two bound states, interacting either through a single $G$-link (term $G_1$) or two $G$-links (term $G_0^2$, with an extra symmetry factor $1/2$, and a flipped sign interpreted as a cancellation of over-counted diagrams).

To compare with the result given in \cite{Bourgine2015}, we shall also absorb the tadpole integral of $Y_0(x)$, i.e. here the last term in the first line. This is done by a finite renormalisation of the coupling constant $q$, $q\to e^{-(1/2)k(0)\e_2}q$, which, at the order of interest, is equivalent to the substitution of $q$ by the original coupling $\L$ in the potential $V(x)=\log q Q(x)$. A different interpretation of this tadpole term was given in \cite{Bourgine2015}, in connexion with the finiteness of the kernel $k(x-y)$ for coinciding points.

The remaining terms now become straightforward. As shown in \cite{Bourgine2015}, the first term in \ref{final} is the logarithm of the Fredholm determinant of (minus) the Hessian matrix associated to the NS limit effective action \cite{Nekrasov2009, Meneghelli2013,Bourgine2014} 
\begin{equation}\label{action_NS}
\CS_0[\rho,\vphi]=\hf\int{\rho(x)\rho(y)G_0(x-y)dxdy}+\int{\rho(x)\vphi(x)dx}+\int{\text{Li}_2(Q_0(x)e^{-\vphi(x)})\dfrac{dx}{2i\pi}},
\end{equation}
with $Q_0(x)=e^{V_0(x)}=qQ(x)+O(\e_2)$ and the classical fields 
$\rho(x)$ and $\vphi(x)$ related to the ($G$-trees) dressed vertices by
\begin{equation}
Y_0(x)=2i\pi\rho(x),\quad \YGz(x)=Q_0(x)e^{-\vphi(x)}.
\end{equation}
Finally, the second term in \ref{final} originates from the dependence of the vertex potential $Q(x)$ on $\e_2$.

In summary, we have presented an alternative derivation for the finite $\e_2$-corrections of the prepotential for $\mathcal{N}=2$ $SU(N_c)$ SYM. This derivation provides the explicit form of the cluster diagrams describing the interactions of instantons, hence complementing the field theory picture drawn in \cite{Bourgine2015}, also for future developments. These Mayer diagrams distinguish long- and short-range interactions, thus shading light on how these two interactions are intertwined. At small $\e_2$, the $p$-clustering (or pinching phenomenon) validates the description in terms of bound states, represented by meta-vertices corresponding to clusters of instantons with only short-range interactions. These bound states lie in an external potential $\log(\L Q(x))$, and interact through a kinetic kernel $k(x)$. In the end it generates the $\text{Li}_2$ term in the classical NS action \ref{action_NS}. At the quantum level (next-to-leading order), the summation of one-loop diagrams, dubbed here $G1$-cycles, produces the usual determinant of the Hessian of the classical action. The simplicity of the final quantum correction relies on several remarkable cancellations, in particular between the corrections to the bound states self-energy (appearing in the $G$-trees clusters) and the hadronic tadpoles (or $G1$-cycles of order one). Thus, most of the finite corrections are given by a quantum treatment of the Nekrasov-Shatashvili action in agreement with the field theory treatment of \cite{Bourgine2015}. The only exception is the integral of $Y_0\nabla Y_0$ in \ref{final} that renders the internal structure of the bound states (like the term $\text{Li}_2$ at the classical level). This term involves a newly introduced operator $\nabla$ that distinguishes the singularities within and outside the integration contour in the instanton moduli space. We may conjecture that this operator shall be relevant also at further orders.

As for a perspective, we would like to come back to the possible connexion of the present partition function with the Operator Product Expansion (OPE) \cite{Anope-BSV} for MHV gluon amplitudes/Wilson loops (WLs) in ${\cal N} = 4$ SYM \cite{AM-amp}. This consideration has been proposed in \cite{FPR2, BFPR} because of the kernel decomposition \ref{Kdecomp} in both schemes provided $\e_2\sim i/\sqrt{\l}$. In particular it relates the weak $\Omega$-background $\epsilon_2 \sim 0$ with the string strong coupling  regime $\lambda \gg 1$, and may facilitate the string one-loop computation of the amplitude/WL. In fact, the WL in the leading strong coupling $ g\sim i/ \epsilon_2 \to +\infty$ becomes an infinite sum over 'mesons' and their bound states so to give rise to TBA-like equations as saddle point equations for a Yang-Yang functional \cite{FPR2, BFPR}. Nevertheless, there is a fundamental difference between the two cases: in the present one the integration contour is closed; while in the WL case, instead, the contour is open, and, in order to become closed, it requires the addition of an additional curve \cite{BFPR}. Moreover, the meson is a composite particle, differently from the instanton, made up of a fermion and an anti-fermion, and  the integrals of unbound fermions over the extra curve contribute at next-to-leading order (nlo), whilst in the present case additional subleading terms are generated by the presence of poles in the external potential \cite{BFPR}.

Eventually, it would be of primary interest to see if the relative agreement with the quantum field theory treatment persists at the next $\e_2$-order, and if additional terms appear reflecting the composite nature of bound states. Since the long-range fluctuation, $X$, has been separated by the  short-range one in our previous field theory argument, the full partition function reduces to the smooth\footnote{At least in the sense that saddle point can be applied systematically for small $\e_1$ and $\e_2$.} $X$-averaging of the short-range partition function $\CZ_\text{short}[X]$ \cite{Bourgine2015}. The latter should permit a systematic ($\e_2$-perturbative) computation. In this context, we really would like to achieve a comparison with the small $\e_1$ and $\e_2$ expansion of \cite{Albe-Nemk} (namely the modular anomaly equation) and the preservation of the S-duality therein.

\medskip
{\bf Acknowledgements}
It is a pleasure to thank A. Lerda and R. J. Szabo for useful discussions. J-E Bourgine thanks I.N.F.N. for his post-doctoral fellowship within the grant GAST, which has also partially supported this project, together with the UniTo-SanPaolo research  grant Nr TO-Call3-2012-0088 {\it ``Modern Applications of String Theory'' (MAST)}, the ESF Network {\it ``Holographic methods for strongly coupled systems'' (HoloGrav)} (09-RNP-092 (PESC)) and the MPNS--COST Action MP1210.

\appendix
\section{Evaluation of $p$-clusters integrals}\label{AppA}
Once the appropriate approximation over the potential has been performed, we are let with cluster integrals involving only $p$-links and a trivial potential. This type of integrals can be evaluated by employing the generalisation of a method originally developed in \cite{Moore1997}, and later used in \cite{Bourgine2014} to calculate $I_l(1)$. It starts with the interpretation of the sum over clusters as the Mayer expansion of nested coupled integrals. Using a formula derived from the Cauchy determinant,
\begin{equation}\label{Cauchy}
(-1)^l\e_2^{-l}\prod_{\superp{i,j=1}{i\neq j}}^l\dfrac{\phi_{ij}}{\phi_{ij}-\e_2}=\sum_{\s\in S_l}{(-1)^\s\prod_{i=1}^l\dfrac1{\phi_i-\phi_{\s(i)}-\e_2}},
\end{equation}
the kernel is expanded as a sum over permutations. In all the cases presented here, only permutations that consists of a single cycle, or a product of two cycles, give a non-zero contribution to the sum. These contributions are then evaluated exactly exploiting the covariance of the integrand under the permutation group.

\subsection{Evaluation of $J_l(1)$}
The simplest integral we have to evaluate is $J_l(\a)$ defined in \ref{def_Jl} at $\a=1$. As for $I_l(\a)$, the sum over clusters in the definition of $J_l(\a)$ is equivalent to the Mayer expansion of the correlation function
\begin{equation}
J_l(\a)=\dfrac1{(l-1)!\e_2^{l}}\int{\prod_{i=1}^{l-1}\dfrac{d\phi_i}{2i\pi}\prod_{\superp{i,j=1}{i<j}}^l\left(1+\e_2p(\phi_{ij})\right)\ \ \sum_{k=1}^{l-1}(\phi_k-x)},
\end{equation}
with the $l$th variable identified with the root, $\phi_l\equiv x$. A priori, the Mayer expansion contains a sum over disconnected clusters. However, contributions of these clusters factorise into connected parts contributions, and those contributions are vanishing if they do not contain a fixed variable, i.e. if the corresponding connected cluster do not contain the root $x$. Thus, only the connected clusters $\D_l^x$ remain in the Mayer expansion \ref{def_Jl}. At $\a=1$, $p(x)$ is equivalent to $p_0(x)$ and the kernel can be expanded on the symmetric group $S_l$ using the Cauchy formula \ref{Cauchy},
\begin{equation}
J_l(1)=\dfrac{(-1)^l}{(l-1)!}\sum_{\s\in S_l}{(-1)^\s\int{\prod_{i=1}^{l-1}\dfrac{d\phi_i}{2i\pi}\prod_{i=1}^l\dfrac1{\phi_i-\phi_{\s(i)}-\e_2}}}\sum_{k=1}^{l-1}(\phi_k-x).
\end{equation}
These nested integrals can always be factorised into a product where factors correspond to the elementary cycles in the decomposition of the permutation $\s$. Due to the pole structure, the integrals are vanishing if the corresponding cycle does not contain the fixed variable $l$. It implies that $\s$ must be a cycle of length $l$, there are $(l-1)!$ such cycles, and they have the signature $(-1)^{l-1}$. Exploiting the invariance under permutations of the integrand, we may change the labels such that $\s(k)=k+1,\ \forall k<l$, $\s(l)=1$. As a result,
\begin{equation}
J_l(1)=-\int{\prod_{i=1}^{l-1}\dfrac{d\phi_i}{2i\pi}\prod_{i=1}^l\dfrac1{\phi_i-\phi_{i+1}-\e_2}\ \sum_{k=1}^{l-1}(\phi_k-x)}.
\end{equation}
Each integral can be evaluated as a sum over residues, leading to
\begin{equation}
J_l(1)=\dfrac{1}{l}\sum_{k=1}^{l-1}(l-k)=\hf(l-1).
\end{equation}
It is possible to further check this result against an explicit computation of the integrals in their original form for small $l$.

\subsection{Computation of $I_l'(1)$}
The previous method works in a similar way, although more involved, to compute $I_l'(1)$. First, notice that the sum over rooted clusters $\D_l^x$  that defines the series of integrals $I_l(\a)$ in \ref{def_Il} appears as the Mayer expansion of
\begin{equation}
I_l(\a)=\dfrac1{(l-1)!\e_2^{l-1}}\int{\prod_{i=1}^{l-1}\dfrac{d\phi_i}{2i\pi}\prod_{\superp{i,j=1}{i<j}}^l\left(1+\e_2p(\phi_{ij})\right)},
\end{equation}
where the last variable $\phi_l=x$ is held fixed. Expanding around $\a=1$, we deduce
\begin{equation}
I_l'(1)=\dfrac1{(l-1)!\e_2^{l-1}}\int{\prod_{i=1}^{l-1}\dfrac{d\phi_i}{2i\pi}\prod_{\superp{i,j=1}{i<j}}^l\dfrac{\phi_{ij}^2}{\phi_{ij}^2-\e_2^2} \sum_{\superp{i,j=1}{i<j}}^l \dfrac{\e_2^2}{\phi_{ij}^2}}.
\end{equation}
The extra poles on the real axis at $\phi_i=\phi_j$ are spurious, but in order to employ the Cauchy formula we need a regularization. Observe that the integral can be obtained as the limit
\begin{equation}
I_l'(1)=\lim_{\g\to0}\dfrac1{(l-1)!\e_2^{l-1}}\int{\prod_{i=1}^{l-1}\dfrac{d\phi_i}{2i\pi}\prod_{\superp{i,j=1}{i<j}}^l\dfrac{\phi_{ij}^2}{\phi_{ij}^2-\e_2^2} \sum_{\superp{i,j=1}{i<j}}^l \dfrac{\e_2^2}{\phi_{ij}^2-\g^2}}.
\end{equation}
with $\g\in i\mathbb{R}^+$. It is now possible to employ the Cauchy determinant formula \ref{Cauchy} to write down
\begin{equation}
I_l'(1)=\lim_{\g\to0}\dfrac{(-1)^l\e_2^3}{(l-1)!}\sum_{\superp{i,j=1}{i<j}}^l\sum_{\s\in S_l}{(-1)^\s\int{\prod_{k=1}^{l-1}\dfrac{d\phi_k}{2i\pi}\prod_{k=1}^l\dfrac1{\phi_k-\phi_{\s(k)}-\e_2}}} \ \dfrac{1}{\phi_{ij}^2-\g^2}.
\end{equation}

The next step is to consider the decomposition of the permutation $\s$ as a product of cycles. At fixed $(i,j)$, three variables single out: $\phi_i$, $\phi_j$ and $\phi_l$. If a cycle contains none of these special points $i$, $j$ or $l$, its contribution factorises, and evaluates to zero. Thus, non-vanishing contributions to the summation over $S_l$ are associated to permutations that decompose into at most three cycles.

Let us first consider the case of exactly three cycles. The contribution of the cycle that contains $\phi_l$ factorises and is non-zero. In the remaining factor, let us assume that the cycle containing $\phi_i$ is of length $k_1$ and the one containing $\phi_j$ of length $k_2$. Without loosing generality (i.e. up to permutations) we can assume that $i=k_1$ and $j=k_1+1$ and $\s(r)=r+1$ within each cycle (up to $\s(k_1)=1$ and $\s(k_1+k_2)=k_1+1$). Thus, the corresponding factor is
\begin{equation}
\int{\prod_{r=1}^{k_1+k_2}\dfrac{d\phi_r}{2i\pi}\prod_{r=1}^{k_1-1}\dfrac1{\phi_r-\phi_{r+1}-\e_2}\dfrac1{\phi_{k_1}-\phi_1-\e_2}\prod_{s=k_1+1}^{k_1+k_2-1}\dfrac1{\phi_s-\phi_{s+1}-\e_2}\dfrac1{\phi_{k_1+k_2}-\phi_{k_1+1}-\e_2}\dfrac1{(\phi_{k_1}-\phi_{k_1+1})^2-\g^2}}.
\end{equation}
The integrals for $\phi_r$ and $\phi_s$ with $r=1\cdots k_1-1$ and $s=k_1+2\cdots k_1+k_2$ contain only a single contributing pole and can be easily evaluated. The evaluation of the residues implies that $\phi_1=\phi_{k_1}+(k_1-1)\e_2$ and $\phi_{k_1+2}=\phi_{k_1+1}+(k_2-1)\e_2$. Only remains an integral over $\phi_{k_1}$ and $\phi_{k_1+1}$,
\begin{equation}
\dfrac1{k_1k_2\e_2^2}\int{\dfrac{d\phi_{k_1}d\phi_{k_1+1}}{(2i\pi)^2}\dfrac1{(\phi_{k_1}-\phi_{k_1+1})^2-\g^2}}.
\end{equation}
Now one of the integrals can be evaluated at the pole, but the second one will be vanishing since there will be no pole left. We conclude that the contributions from permutations that decompose into a product of three cycles are vanishing, and
\begin{equation}\label{Ilp}
I_l'(1)=\dfrac{(-1)^l\e_2^3}{(l-1)!}\lim_{\g\to0}(I_\text{2-cycles}(\g)+I_\text{1-cycles}(\g)).
\end{equation}

Next, we consider permutations that decompose into a product of two cycles. If both $i$ and $j$ belong to the first cycle, and $l$ to the second cycle, a similar argument than in the case of three cycles holds: the contribution of the second cycle factorises and we find for the first cycle of length $k$, taking $i=1$,
\begin{equation}
\int{\prod_{r=1}^k\dfrac{d\phi_r}{2i\pi}\prod_{r=1}^{k-1}\dfrac1{\phi_{r}-\phi_{r+1}-\e_2}\dfrac1{\phi_k-\phi_1-\e_2}\dfrac1{(\phi_1-\phi_j)^2-\g^2}}.
\end{equation}
Evaluating all the integrals except for $\phi_1$ and $\phi_j$ gives $\phi_2=\phi_j+(j-2)\e_2$ and $\phi_{j+1}=\phi_1+(k-j)\e_2$, and
\begin{equation}
\int{\dfrac{d\phi_1d\phi_j}{(2i\pi)^2}\dfrac1{\phi_1-\phi_j-(j-1)\e_2}\dfrac1{\phi_j-\phi_1-(k-j+1)\e_2}\dfrac1{(\phi_1-\phi_j)^2-\g^2}}.
\end{equation}
Further computing the sum of residues for $\phi_1$, we obtain an expression that does not depend on $\phi_j$, and the total contribution vanishes. Thus, non-vanishing contributions should have one of the two variables $i$ or $j$ in the cycle containing $l$. Keeping this fact in mind, we decompose $\s$ into a product of two cycles, one of length $k$, the second of length $l-k$ with the latter containing the fixed variable $l$. There are $C_{l-1}^{k}\times (k-1)!\times (l-k-1)!=(l-1)!/k$ such permutations, with total signature $(-1)^{k-1+l-k-1}=(-1)^l$. Using the invariance of the integrand under permutations, it is possible to assume that the first cycle is $(1\ 2\ \cdots \ k)$ and the second cycle $(k+1\ k+2\ \cdots \ l)$, so that
\begin{align}
\begin{split}
I_\text{2-cycles}(\g)=(-1)^l(l-1)!\sum_{k=1}^{l-1}\dfrac1k\int\prod_{r=1}^{l-1}\dfrac{d\phi_r}{2i\pi}&\prod_{r=1}^{k-1}\dfrac1{\phi_{r}-\phi_{r+1}-\e_2}\dfrac1{\phi_k-\phi_1-\e_2}\\
&\times\prod_{s=k+1}^{l-1}\dfrac1{\phi_{s}-\phi_{s+1}-\e_2}\dfrac1{\phi_l-\phi_{k+1}-\e_2}\ \sum_{\superp{i,j=1}{i<j}}^l\dfrac1{\phi_{ij}^2-\g^2}.
\end{split}
\end{align}
We have shown previously that the permutations such that $i<j\leq k$ (both $i$ and $j$ in the first cycle) or such that $k+1\leq i<j$ (both $i$ and $j$ in the second cycle) produce a vanishing contribution. Thus, the only remaining case to treat is $1\leq i\leq k$ and $k+1\leq j\leq l$. The integrals over $\phi_r$ with $r\neq i,j$ contain only a single pole in the upper half plane. The evaluation as a sum over the corresponding residues brings the constraints $\phi_{i+1}=\phi_i+(k-1)\e_2$, $\phi_{j+1}=\phi_j+(l-j-1)\e_2$ and $\phi_{k+1}=\phi_j+(j-k-1)\e_2$, which gives
\begin{equation}
I_\text{2-cycles}(\g)=(-1)^l(l-1)!\sum_{k=1}^{l-1}\dfrac1k\sum_{i=1}^k\sum_{j=k+1}^l\times -\dfrac1{k\e_2}\int{\dfrac{d\phi_id\phi_j}{(2i\pi)^2}\dfrac1{\phi_j-\phi_l-(l-j)\e_2}\dfrac1{\phi_l-\phi_j-(j-k)\e_2}\dfrac1{\phi_{ij}^2-\g^2}}.
\end{equation}
Further evaluating the remaining integrals produces the $\g$-diverging piece
\begin{equation}
I_\text{2-cycles}(\g)=(-1)^l\dfrac{(l-1)!}{2\g\e_2^2}\sum_{k=1}^{l-1}\dfrac1k.
\end{equation}

It only remains to treat the contributions of single cycles. There are $(l-1)!$ such permutations and their signature is $(-1)^{l-1}$. We can use the invariance under permutations to set $\s(i)=i+1$, $\s(l)=1$, and write
\begin{equation}
I_\text{1-cycles}(\g)=(-1)^{l-1}(l-1)!\sum_{\superp{i,j=1}{i<j}}^{l}\int{\prod_{r=1}^{l-1}\dfrac{d\phi_r}{2i\pi}\prod_{r=1}^{l-1}\dfrac1{\phi_r-\phi_{r+1}-\e_2}\dfrac1{\phi_l-\phi_1-\e_2}\dfrac1{(\phi_i-\phi_j)^2-\g^2}}.
\end{equation}
Again, we evaluate all the integrals except $\phi_i$ and $\phi_j$: $\phi_1=\phi_i+(i-1)\e_2$, $\phi_{j+1}=\phi_l+(l-j-1)\e_2$ and $\phi_{i+1}=\phi_j+(j-i-1)\e_2$:
\begin{equation}
I_\text{1-cycles}(\g)=(-1)^{l-1}(l-1)!\sum_{\superp{i,j=1}{i<j}}^{l}\int{\dfrac{d\phi_id\phi_j}{(2i\pi)^2}\dfrac1{\phi_i-\phi_j-(j-i)\e_2}\dfrac1{\phi_l-\phi_i-i\e_2}\dfrac1{\phi_j-\phi_l-(l-j)\e_2}\dfrac1{(\phi_i-\phi_j)^2-\g^2}}.
\end{equation}
Evaluating the last integrals, we get
\begin{equation}
I_\text{1-cycles}(\g)=(-1)^{l-1}(l-1)!\sum_{\superp{i,j=1}{i<j}}^{l}\dfrac{l\e_2+2\g}{2\e_2\g l(\g+(j-i)\e_2)(\g+(l-j+i)\e_2)}.
\end{equation}
The summand only depends on the difference $k=j-i$. As $j=1\cdots l$ and $i=1\cdots j-1$, we have $k=1\cdots l-1$ with the $k^{\text{th}}$ term repeating $(l-k)$ times,
\begin{equation}
I_\text{1-cycles}(\g)=(-1)^{l-1}(l-1)!\sum_{k=1}^{l-1}\dfrac{(l\e_2+2\g)(l-k)}{2\e_2\g l(\g+k\e_2)(\g+(l-k)\e_2)}=(-1)^{l-1}\dfrac{(l-1)!}{2\e_2\g}\sum_{k=1}^{l-1}\dfrac1{\g+k\e_2}.
\end{equation}
Expanding in $\g\to0$,
\begin{equation}
I_\text{1-cycles}(\g)=(-1)^{l-1}\dfrac{(l-1)!}{2\e_2\g}\sum_{k=1}^{l-1}\left(\dfrac1{k\e_2}-\dfrac{\g}{(k\e_2)^2}+O(\g^2)\right).
\end{equation}
Combining one and two cycles contributions in \ref{Ilp} to derive $I_l'(1)$, we find that the terms of order $O(1/\g)$ cancel each others and a term of order one remains,
\begin{equation}
I_l'(1)=\hf\sum_{k=1}^{l-1}\dfrac1{k^2}.
\end{equation}
This expression can be checked for small values of $l$ against the computation of the original integral formula. It is easy to recover the first values $I_2'(1)=1/2$, $I'_3(1)=5/8$ and $I_4'(1)=49/72$.

\subsection{Bi-rooted integrals}\label{AppB}
The integral $\Ib_l^{x,y}$ is defined as a sum over a set of bi-rooted $p$-clusters $\D_l^{x,y}$ of $l$ vertices in which the direct connection of the roots $x$ and $y$ is prevented,
\begin{equation}
\Ib_l^{x,y}=\e_2^{-(l-2)}\sum_{\bar\D_l^{x,y}}\dfrac1{\s(\bar\D_l^{x,y})}\int{\prod_{\superp{i\in V(\bar\D_l^{x,y})}{x\neq i\neq y}}\dfrac{d\phi_i}{2i\pi}\prod_{<ij>\in E_p(\bar\D_l^{x,y})}\e_2 p_0(\phi_i-\phi_j)}.
\end{equation}
This integral is related to the sum over unrestricted bi-rooted $p$-clusters $\D_l^{x,y}$ defined as
\begin{equation}\label{def_Il_xy}
I_l^{x,y}=\e_2^{-(l-2)}\sum_{\D_l^{x,y}}\dfrac1{\s(\D_l^{x,y})}\int{\prod_{\superp{i\in V(\D_l^{x,y})}{x\neq i\neq y}}\dfrac{d\phi_i}{2i\pi}\prod_{<ij>\in E_p(\D_l^{x,y})}\e_2 p_0(\phi_i-\phi_j)}
\end{equation}
as follows. Consider an unrestricted cluster of the type $\D_l^{x,y}$. If the roots are not connected with a $p$-link, this cluster is also of the type $\bar\D_l^{x,y}$. Otherwise, if the $p$-link connecting the roots is not the only path between them, we recover a cluster of type $\bar \D_l^{x,y}$ by removing this link. The last configuration is the case of roots only connected through a single $p$-link and no other path. Then, removing the $p$-link produces two disconnected rooted clusters $\D_k^x$ and $\D_{l-k}^y$ with $1\leq k<l$. Thus, we formally have the bijection
\begin{equation}
\{\D_l^{x,y}\}\equiv\{\bar\D_l^{x,y}\}\cup\{p\bar\D_l^{x,y}\}\cup\sum_{k=1}^{l-1}\{\D_k^xp\D_{l-k}^y\},
\end{equation}
which provide the following relation between the two integrals
\begin{equation}\label{rel_I_Ib}
I_l^{x,y}=(1+\e_2p_0(x-y))\Ib_l^{x,y}+\e_2p_0(x-y)\d_l,\quad \d_l=\sum_{k=1}^{l-1}I_k(1)I_{l-k}(1)=\dfrac2l\sum_{k=1}^{l-1}\dfrac1k.
\end{equation}
Note that there cannot be enhancement of the symmetry in the last term due to the fact that we have bi-rooted clusters and roots cannot be exchanged: $\s(\D_k^xp\D_{l-k}^y)=\s(\D_k^x)\s(\D_{l-k}^y)$. The integrals $I_k(1)$ and $I_{l-k}(1)$ are associated to the rooted clusters $\D_k^x$ and $\D_{l-k}^y$ respectively. It is a remarkable fact that they do not depend on the root variables $x$ or $y$, and as a consequence they yield to zero when integrated with respect to these variables.

The sum of the integrals over the clusters $\D_l^{x,y}$ in the definition \ref{def_Il_xy} of $I_l^{x,y}$ can be interpreted as the connected terms in the Mayer cluster expansion of
\begin{equation}
\tilde{I}_l^{x,y}=\dfrac{\e_2^{2-l}}{(l-2)!}\int{\prod_{i=2}^{l-1}\dfrac{d\phi_i}{2i\pi}\prod_{\superp{i,j=1}{i<j}}^l\left(1+\e_2p_0(\phi_i-\phi_j)\right)},
\end{equation}
with the two variables $\phi_1=x$ and $\phi_l=y$ fixed. This expansion is over disconnected clusters, but these clusters will be vanishing if the roots do not belong to them. Thus, the series contains only terms associated to a single cluster containing both roots (contributions to $I_l^{x,y}$), or to two clusters each with one root (reproducing $\d_l$), and combining with \ref{rel_I_Ib},
\begin{equation}\label{rel_It_Ib}
\tilde{I}_l^{x,y}=I_l^{x,y}+\d_l=(1+\e_2p_0(x-y))(\Ib_l^{x,y}+\d_l).
\end{equation}

We now address the issue of computing $\tilde{I}_l^{x,y}$. The formula \ref{Cauchy} provides an expansion of the kernel over the sum of permutations in $S_l$,
\begin{equation}
\tilde{I}_l^{x,y}=\dfrac{(-1)^l\e_2^2}{(l-2)!}\sum_{\s\in S_l}(-1)^\s\int{\prod_{i=2}^{l-1}\dfrac{d\phi_i}{2i\pi}\prod_{i=1}^l\dfrac1{\phi_i-\phi_{\s(i)}-\e_2}},
\end{equation}
withe the identification of fixed variables $\phi_1=x$ and $\phi_l=y$. Again, permutations can be decomposed into a product of cycles, and the integrations over variables in different cycles decouple, thus factorising the initial integral into factors in correspondence with the cycles in the decomposition. If neither $1$ or $l$ belongs to the cycle, the associated integral vanishes. This gives only two types of contributions to the summation. The first contribution comes from cycles of length $l$ that contain both $1$ and $l$. Using the invariance under permutations of the integrand, we can set $\s(i)=i+1$ and $\s(l)=1$, at the cost of loosing the information on the root position, with now $\phi_k=x$ with $k=1\cdots l-1$. There are $(l-1)!$ cycles of length $l$, but only $(l-2)!$ for each fixed $k$. Their signature is $(-1)^{l-1}$, and the contribution reads
\begin{equation}
-\e_2^2\sum_{k=1}^{l-1}\int{\prod_{\superp{i=1}{i\neq k}}^{l-1}\dfrac{d\phi_i}{2i\pi}\prod_{i=1}^{l-1}\dfrac1{\phi_i-\phi_{i+1}-\e_2}\dfrac1{\phi_l-\phi_1-\e_2}}.
\end{equation}
Each integration variable has only a single pole in the upper half plane, at $\phi_i=\phi_{i+1}+\e_2$. Evaluating the integrals over residues, we find by induction $\phi_1=x+(k-1)\e_2$, $\phi_{k+1}=y+(l-k-1)\e_2$ giving the contribution
\begin{equation}
-\e_2^2\sum_{k=1}^{l-1}\dfrac1{x-y-(l-k)\e_2}\dfrac1{y-x-k\e_2}=\dfrac2l\sum_{k=1}^{l-1}\dfrac{k\e_2^2}{(x-y)^2-(k\e_2)^2}.
\end{equation}
The second contribution corresponds to permutations that factorise into a product of two cycles, the first one of length $k$ should contain $1$ and the second one, with length $l-k$, contains $l$. There are $C_{l-2}^{k-1}\times(k-1)!\times(l-k-1)!=(l-2)!$ such permutations at fixed $k=1\cdots l-1$. The initial integral factorises into a product associated to each cycle. Within the cycles, we can order the integration variables such that $\s(i)=i+1$ and after evaluation of the residues, we find the contribution
\begin{equation}
\dfrac{(-1)^l\e_2^2}{(l-2)!}\sum_{k=1}^{l-1}(-1)^{k-1+l-k-1}(l-2)!\times\dfrac{-1}{k\e_2}\times\dfrac{-1}{(l-k)\e_2}=\dfrac2l\sum_{k=1}^{l-1}\dfrac1k,
\end{equation}
which coincide with $\d_l$. Re-assembling the two contributions, we get
\begin{equation}\label{res_Ilxy}
\tilde{I}_l^{x,y}=\dfrac2l\sum_{k=1}^{l-1}\dfrac{(x-y)^2}{k\left[(x-y)^2-(k\e_2)^2\right]},\quad I_l^{x,y}=\dfrac2l\sum_{k=1}^{l-1}\dfrac{k\e_2^2}{(x-y)^2-(k\e_2)^2}.
\end{equation}
It is easy to verify that
\begin{equation}
\int\dfrac{dy}{2i\pi}I_l^{x,y}=\int\dfrac{dy}{2i\pi}\tilde{I}_l^{x,y}=\dfrac{l-1}{l}\e_2=(l-1)I_l(1)\e_2,
\end{equation}
which is also a consequence of the combinatorial identity (B.6) in \cite{Bourgine2013} between sum over rooted and bi-rooted clusters.

Inverting the relation \ref{rel_It_Ib}, we deduce
\begin{equation}\label{int_Ib}
\Ib_l^{x,y}=\dfrac2l\sum_{k=1}^{l-1}\dfrac{(k^2-1)\e_2^2}{k\left[(x-y)^2-(k\e_2)^2\right]}, \quad\text{and}\quad 
\int{\dfrac{dy}{2i\pi}\Ib_l^{x,y}}=\dfrac{\e_2}{l}\left[l-1-\sum_{k=1}^{l-1}\dfrac1{k^2}\right].
\end{equation}

\section{Necklace and irreducible $p$-clusters}\label{AppC}
\begin{figure}[!t]
\centering
\includegraphics[width=9cm]{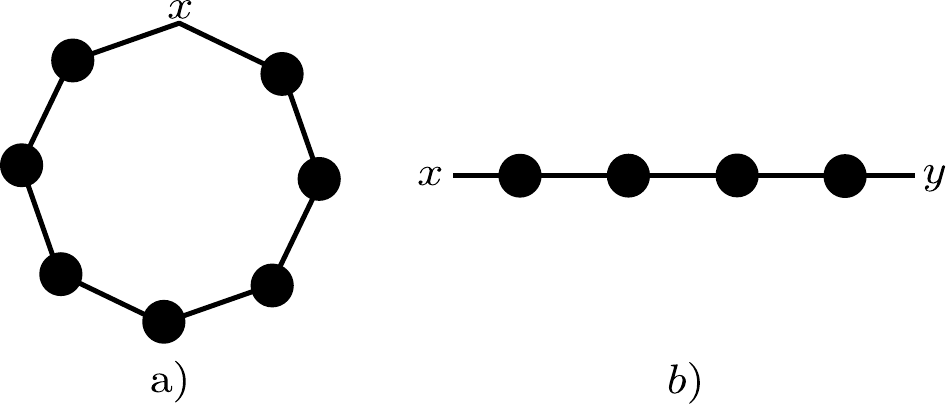}
\caption{a) A rooted necklace with eight vertices and root $x$. b) A bi-rooted chain of six vertices and roots $x,y$.}
\label{necklace}
\end{figure}

To justify the assumption that only articulation links provide $\e_2$-corrections to $\CZ_\text{short}$, it has been claimed in \cite{Bourgine2015} that irreducible clusters, which by definition do not contain any articulation link, have a vanishing term at the order $O(\e_2)$ in the $\e_2$-expansion. In this appendix, we show that this is indeed the case for necklace diagrams represented in figure \refOld{necklace}$a$. As in \cite{Bourgine2015}, to each vertex of the $p$-clusters $\D_l$ is associated a potential $U(x)$ independent of $\e_2$, and to each link $<ij>$ the interaction $\e_2p_0(\phi_{ij})$.\footnote{Possible $\e_2$-dependences of the potential $U(x)$ can be treated separately.} Our argument does not depend on the symmetry coefficients but lies at the level of the nested integrations. It is thus possible, and it will also be more convenient, to work with rooted clusters, the remaining integration over the root being left as the last step to be performed.

First, we introduce the contribution of a bi-rooted chain with the roots $x$ and $y$, and $n-2$ intermediate vertices (see figure \refOld{necklace}$b$),
\begin{equation}
\CC_n(x,y)=\int{\prod_{i=2}^{n-1}U(\phi_i)\dfrac{d\phi_i}{2i\pi}\ p_0(x-\phi_2)\prod_{i=2}^{n-2}p_0(\phi_i-\phi_{i+1})\ p_0(\phi_{n-1}-y)}.
\end{equation}
To simplify some expressions, we have decided not to include the potential of the roots in the definition, so that $\CC_2(x,y)=p_0(x-y)$. These quantities are obviously symmetric in the exchange of the roots $x\leftrightarrow y$. Note also that the $n-1$ links bring a factor $\e_2^{n-1}$ and the $n$ vertices a factor $\e_2^{-n}$. Because of the additional $\e_2$ factor in the definition of the prepotential, there are no $\e_2$ variables left in the expression of $\CC_n(x,y)$. These quantities can also be constructed recursively by adding to the necklace of order $n$, $\CC_n(\phi,y)$ a new link $<x\phi>$ and then integrating of the vertex $\phi$ with the appropriate potential $U(\phi)$ inserted,
\begin{equation}\label{rec_cn}
\CC_{n+1}(x,y)=\int{p_0(x-\phi)\CC_n(\phi_,y)U(\phi)\dfrac{d\phi}{2i\pi}}=\hf\left[U(x)\CC_n(x,y)\right]^{+_x}.
\end{equation}
In order to disambiguate the notation, we have added up subscript $x$ to the operation $[\cdots]^+$ defined in \ref{def_plus} when it is acting on the variable $x$. The contribution attached to the rooted necklace can be obtained from the rooted chain by inserting a link between the two roots, and then integrating over one of the root (after insertion of the corresponding potentials):
\begin{equation}
\CC_n(x)=U(x)\int{\e_2p_0(x-y)\CC_n(x,y)U(y)\dfrac{dy}{2i\pi}}.
\end{equation}
In this expression of $\CC_n(x)$ we have included the potential $U(x)$ of the root, in contrast with the definition of $\CC_n(x,y)$. As a consequence of the recursive construction \ref{rec_cn} of the latter, we have $\CC_n(x)=\e_2U(x)\CC_{n+1}(x,x)$.

In order to understand the general argument, it is simpler to start from the case of order $n=2$, although this cluster is non-physical (i.e. it does not appear in the Mayer cluster expansion since it consists of two vertices attached by two links). Instead of performing a direct evaluation of $\CC_2(x)$, it is instructive to first compute the intermediate quantity 
\begin{equation}
\CC_3(x,y)=\dfrac14 p_0(x-y)^{+_x}\left[U(x)^{+_x}+U(y)^{+_y}-2\e_2\dfrac{\bar U(x)^{+_x}-\bar U(y)^{+_y}}{x-y}\right],
\end{equation}
with $\bar U(x)=U_\text{reg.}(x)-U_\text{sing.}(x)$. This expression has been obtained by evaluating the $p_0$ convolution of $U(x)\CC_2(x,y)$. In this integration, and all the similar ones given below, the potential is treated as follows. First we assume $U(x)=1/(x-r)$ with $r$ either inside or outside the integration contour (both situations are treated separately). We rewrite the result as a linear function of this potential. Since the result is linear in the potential, it is possible to act with $\p_r$ to obtain higher order poles. It is also possible to take a linear combination of various poles, leading to a result for any rational function $U(x)$. The resulting expression for $\CC_3(x,y)$ can be $\e_2$-expanded to produce
\begin{equation}
\CC_3(x,y)=\dfrac14 p_0(x-y)^{+_x}\left[U(x)+U(y)+\e_2\p_x\bar U(x)+\e_2\p_y\bar U(y)-2\e_2\dfrac{\bar U(x)-\bar U(y)}{x-y}+O(\e_2^2)\right],
\end{equation}
with $\p_x \bar U(x)=\nabla U(x)$. Observe that the prefactor $p_0(x-y)^{+_x}$ should not be $\e_2$-expanded due to the poles pinching the integration contour at $x-y\propto\e_2$. Taking the limit $y\to x$, the third term inside the brackets becomes a derivative of $\bar U(x)$ which cancel the two other terms of order $\e_2$, and it only remains
\begin{equation}\label{CC2}
\CC_2(x)=-\dfrac14U(x)+O(\e_2^2).
\end{equation}
In this expression, the term of order $O(\e_2)$ is absent.

To treat the general case, we notice that at each order the poles that pinch the real axis can be factorised,
\begin{equation}\label{CCn}
\CC_n(x,y)=p_0^{[n-2]}(x-y)R_n(x,y),\quad\text{with}\quad p_0^{[n-1]}(x)=\dfrac{n\e_2}{x^2-n^2\e_2^2}=p_0^{[n-2]}(x)^+,
\end{equation}
where $R_n(x,y)$ can now be safely expanded in $\e_2$. This decomposition is proved by induction using the recursive procedure \ref{rec_cn},
\begin{equation}
\CC_{n+1}(x,y)=\int{p_0(x-\phi)p_0^{[n-2]}(\phi-y)U_n(\phi)\dfrac{d\phi}{2i\pi}},\quad\text{with}\quad U_n(\phi)=U(\phi)R_n(\phi,y).
\end{equation}
As the notation suggests, the variable $y$ of $R_n(\phi,y)$ is a spectator in the whole procedure. Then, the potential $U_n(\phi)$ can be treated like any rational potential, in particular it decomposes into a sum of regular and singular parts. Once the integral has been computed, one recovers the form \ref{CCn} with\
\begin{align}
\begin{split}
2nR_{n+1}(x,y)&=(n-1)U_n(x)^++U_n^{[n-1]}(y)-\dfrac{2\e_2(n-1)}{x-y-(n-2)\e_2}\left[U_{n,\text{reg.}}(x+\e_2)-U_{n,\text{reg.}}(y+(n-1)\e_2)\right]\\
&+\dfrac{2\e_2(n-1)}{x-y+(n-2)\e_2}\left[U_{n,\text{sing.}}(x-\e_2)-U_{n,\text{sing.}}(y-(n-1)\e_2)\right],
\end{split}
\end{align}
where we have employed the notation $f^{[n]}(x)=f_\text{reg.}(x+n\e_2)+f_\text{sing.}(x-n\e_2)$ for the $n$th iteration of the operation $[\cdots]^+$. The last two terms can be safely expanded in $\e_2$, even at small distances $x-y\propto\e_2$, since the brackets are vanishing at $y=x\pm (n-2)\e_2$ thus cancelling the spurious poles. At subleading order in $\e_2$, we find
\begin{equation}
2nR_{n+1}(x,y)=(n-1)U_n(x)+U_n(y)+(n-1)\e_2\left[\p_x\bar U_n(x)+\p_y\bar U_n(y)-2\dfrac{\bar U_n(x)-\bar U_n(y)}{x-y}\right]+O(\e_2^2).
\end{equation}
As in the $n=2$ case, the term of order $O(\e_2)$ disappears in the limit $y\to x$ ,
\begin{equation}
R_{n+1}(x,x)=\hf U_n(x)+O(\e_2^2)=\hf U(x) R_n(x,x)+O(\e_2^2).
\end{equation}
By induction, we deduce
\begin{equation}
R_{n+1}(x,x)=\left(\dfrac{U(x)}2\right)^{n-1}+O(\e_2^2),\quad\text{and}\quad \CC_n(x)=-\dfrac{2}{n}\left(\dfrac{U(x)}2\right)^n+O(\e_2^2).
\end{equation}
Thus, there are no contributions of order $O(\e_2)$ from the necklace graphs.

\begin{figure}[!t]
\centering
\includegraphics[width=7cm]{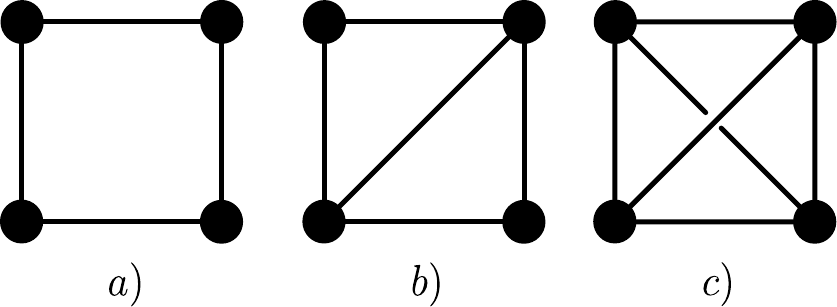}
\caption{Irreducible $p$-clusters with four vertices (symmetry factors: $1/8$, $1/4$, $1/24$).}
\label{square}
\end{figure}
To further test the conjecture that irreducible graphs do not contribute at the order $O(\e_2)$, we have computed all the irreducible graphs with four vertices in the case where the potential is the leading order of the $U(1)$ pure SYM potential,
\begin{equation}
U(x)=\dfrac{q}{(x-a)(x-a+\e_1)}.
\end{equation}
The relevant clusters are displayed in figure \refOld{square}, and their contributions read (without the symmetry factor)
\begin{equation}
a)=\dfrac{5q^4}{8\e_1^7}+O(\e_2^2),\quad b)=-\dfrac{25q^4}{72\e_1^7}+O(\e_2^2),\quad c)=\dfrac{5q^4}{24\e_1^7}+O(\e_2^2).
\end{equation}
These results support our conjecture.

\section{More details on the study of $U(1)$ $\mathcal{N}=2$ SYM}
\begin{figure}[!t]
\centering
\includegraphics[width=6cm]{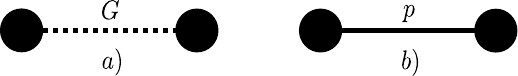}
\caption{The two possible clusters with only two vertices, with $p$-link in plain and $G$-link dashed. Both have a symmetry factor $1/2$.}
\label{2nodes}
\end{figure}
The Nekrasov instanton partition function of $U(1)$ SYM without matter is particularly simple,
\begin{equation}
\CZ_{U(1)}=\exp\left(\dfrac{\Lambda}{\e_1\e_2}\right).
\end{equation}
However, the evaluation of individual clusters exhibits a non-trivial dependence in the gauge coupling, which is remarkably canceled in order to provide the simple result given above. This theory is thus a very good candidate to test our formula \ref{final} for the subleading corrections in $\e_2$ to the prepotential. Here we perform this verification up to the order $O(q^4)$, i.e. four vertices, which already involves up to three $G$-cycles. In particular, this is a good test of the non-contribution of irreducible $p$-clusters at the origin of our conjecture for the $G$-trees corrections. 

In order to simplify our calculations, we work with $q$ fixed instead of $\Lambda$, and will later replace it with the proper gauge theory coupling $\Lambda$. Then, the potential associated to vertices for $U(1)$ SYM expands as
\begin{equation}
Q(x)=\dfrac1{(x-a)(x-a+\e_1+\e_2)}\implies e^{V_0(x)}=\dfrac{q}{(x-a)(x-a+\e_1)},\quad V_1(x)=-\dfrac1{x-a+\e_1}.
\end{equation}
Note that the Coulomb branch vev $a$ can be set to zero due to an invariance under translations. However, we keep it non-zero here mostly for a matter of aesthetics. The function $Y_0(x)$ is obtained by solving perturbatively in $q$ the Nekrasov-Shatashvili NLIE \ref{TBA_Y} for $Y_G^{(0)}$,
\small
\begin{align}
\begin{split}
Y_G^{(0)}(x)=e^{V_0(x)}\Big[&1-\dfrac{3q}{(x-a-\e_1)(x-a+2\e_1)}+\dfrac{10q^2}{(x-a-\e_1)(x-a-2\e_1)(x-a+2\e_1)(x-a+3\e_1)}\\
&-\dfrac{35q^3}{(x-a-3\e_1)(x-a-2\e_1)(x-a-\e_1)(x-a+2\e_1)(x-a+3\e_1)(x-a+4\e_1)}\Big]+O(q^5),
\end{split}
\end{align}
\normalsize
Plugging this solution into \ref{rel_Y0_YG0} we deduce $Y_0(x)$,
\small
\begin{align}
\begin{split}
Y_0(x)&=e^{V_0(x)}-\dfrac{5 \hx^2+5\e_1\hx+2\e_1^2}{2(\hx-\e_1)(\hx+2\e_1)}e^{2V_0(x)}+\dfrac{2(11\hx^4+22\e_1\hx^3+34\e_1^2\hx^2+23\e_1^3\hx+6\e_1^4)}{3(\hx-2\e_1)(\hx-\e_1)(\hx+2\e_1)(\hx+3\e_1)}e^{3V_0(x)}\\
&-\dfrac{93\hx^8+372\e_1\hx^7+944\e_1^2\hx^6+1530\e_1^3x^5+143\e_1^4\hx^4-1830\e_1^5\hx^3-2476\e_1^6\hx^2-1368\e_1^7\hx-288\e_1^8}{4(\hx-3\e_1)(\hx-2\e_1)(\hx-\e_1)^2(\hx+2\e_1)^2(\hx+3\e_1)(\hx+4\e_1)}e^{4V_0(x)}+O(q^5),
\end{split}
\end{align}
\normalsize
with the shortcut notation $\hx=x-a$. Isolating the contributions of the singularities at $x=a,a+\e_1,a+2\e_1,a+3\e_1$ within the integration contour, it is possible to compute $\nabla Y_0(x)$,
\tiny
\begin{align}
\begin{split}
&\nabla Y_0(x)=\dfrac{2\hx^2+2\e_1\hx+\e_1^2}{\hx(\hx+\e_1)\e_1}e^{V_0(x)}-\dfrac{3\hx^6+9\e_1\hx^5+18\e_1^2\hx^4+21\e_1^3\hx^3-\e_1^4\hx^2-10\e_1^5\hx-4\e_1^6}{\hx(\hx-\e_1)^2(\hx+\e_1)(\hx+2\e_1)^2\e_1}e^{2V_0(x)}\\
&+2\dfrac{10\hx^{10}+50\e_1\hx^9+209\e_1^2\hx^8+536\e_1^3\hx^7-16\e_1^4\hx^6-1714\e_1^5\hx^5-2153\e_1^6\hx^4-924\e_1^7\hx^3+690\e_1^8\hx^2+792\e_1^9\hx+216\e_1^10}{3\hx(\hx-2\e_1)^2(\hx-\e_1)^2(\hx+\e_1)(\hx+2\e_1)^2(\hx+3\e_1)^2\e_1}e^{3V_0(x)}+O(q^4).
\end{split}
\end{align}
\normalsize

\begin{figure}[!t]
\centering
\includegraphics[width=10cm]{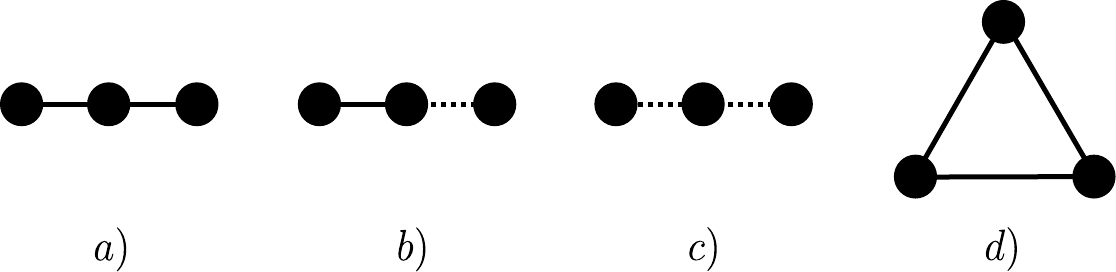}
\caption{The $G$-trees clusters with three vertices which contribute to $\CF_A^{(1)}$. The symmetry factors are respectively $1/2$, $1$, $1/2$ and $1/6$.}
\label{3nodesA}
\end{figure}

To emphasise the various cancellations, we compute $G$-trees and $G1$-cycles terms separately. The contributions to the $G$-trees corrections read
\begin{align}
\begin{split}
&\int{\dfrac{dx}{2i\pi}V_1(x)Y_0(x)}=-\dfrac{q}{\e_1^2}+\dfrac{q^2}{4\e_1^4}-\dfrac{q^3}{9\e_1^6}+\dfrac{11}{192}\dfrac{q^4}{\e_1^8}+O(q^5)\\
&\dfrac{2}{\e_1}\int{\dfrac{dx}{2i\pi}\d L_1\left(1-e^{-Y_0(x)}\right)}=-\dfrac{q^2}{\e_1^4}-\dfrac{q^4}{144\e_1^8}+O(q^5)\\
&\dfrac14\int{\dfrac{dx}{2i\pi}Y_0(x)\nabla Y_0(x)}=\dfrac{q^2}{2\e_1^4}-\dfrac{5q^3}{8\e_1^6}+\dfrac{22q^4}{27\e_1^8}+O(q^5)\\
&\hf \int{\dfrac{dxdy}{(2i\pi)^2}Y_0(x)Y_0(y)G_1(x-y)}=\dfrac{q^2}{4\e_1^4}-\dfrac{q^3}{24\e_1^6}+\dfrac{q^4}{72\e_1^8}+O(q^5).
\end{split}
\end{align}
Summing up all these terms, we find
\begin{equation}\label{CFA}
\CF_A^{(1)}=-\dfrac{q}{\e_1^2}-\dfrac{7q^3}{9\e_1^6}+\dfrac{1519q^4}{1728\e_1^8}+O(q^5).
\end{equation}
We note that there is no contribution at the order $O(q^2)$, this is due to the cancellation between the two diagrams displayed in the figure \refOld{2nodes},
\begin{equation}
a)=\dfrac{q^2}{2\e_1^3}-\dfrac{q^2}{\e_1^4}\e_2+O(\e_2^2),\quad b)=-\dfrac{q^2}{2\e_1^3}+\dfrac{q^2}{\e_1^4}\e_2+O(\e_2^2).
\end{equation}
The term of order $O(q^3)$ in $\CF_A^{(1)}$ is the sum of the four diagrams displayed in the figure \refOld{3nodesA}:
\begin{align}
\begin{split}
&a)=\dfrac{3q^3}{4\e_1^5}-\dfrac{11q^3}{4\e_1^6}\e_2+O(\e_2^2),\quad b)=-\dfrac{5q^3}{4\e_1^5}+\dfrac{11q^3}{4\e_1^6}\e_2+O(\e_2^2),\\
&c)=\dfrac{7q^3}{12\e_1^5}-\dfrac{61q^3}{36\e_1^6}\e_2+O(\e_2^2),\quad d)=-\dfrac{q^3}{12\e_1^5}-\dfrac{q^3}{12\e_1^6}\e_2+O(\e_2^2).
\end{split}
\end{align}
Note that the terms of order $O(1)$ in the $\e_2$-expansion cancel, giving no contribution to $\CF^{(0)}$.

\begin{figure}[!t]
\centering
\includegraphics[width=8cm]{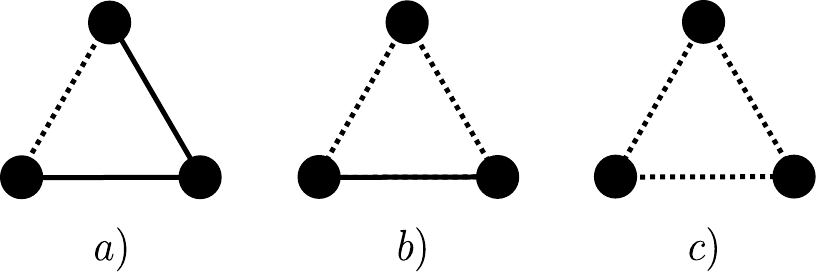}
\caption{The $G1$-cycle clusters with three vertices which contribute to $\CF_B^{(1)}$. The symmetry factors are respectively $1/2$, $1/2$ and $1/6$.}
\label{3nodesB}
\end{figure}

To derive the $G1$-cycles corrections, we expand on the order of the clusters and evaluate
\begin{align}
\begin{split}
&G_0(0)\int{\dfrac{dx}{2i\pi}\left[\hf \D_0(x)-\hf Y_0(x)-\d L_1\left(1-e^{-Y_0(x)}\right)\right]}=\dfrac{3q^3}{2\e_1^6}-\dfrac{343q^4}{144\e_1^8}+O(q^5),\\
&\dfrac14\int{\dfrac{dxdy}{(2i\pi)^2}\left[\D_0(x)\D_0(y)-Y_0(x)Y_0(y)\right]G_0(x-y)^2}=-\dfrac{q^3}{\e_1^6}+\dfrac{4385q^4}{1728\e_1^8}+O(q^5)\\
&\dfrac16\int{\dfrac{dxdydz}{(2i\pi)^3}\D_0(x)\D_0(y)\D_0(z)G_0(x-y)G_0(y-z)G_0(x-z)}=\dfrac{5q^3}{18\e_1^6}-\dfrac{23q^4}{18\e_1^8}+O(q^5)\\
&\dfrac18\int{\dfrac{d\phi_1d\phi_2d\phi_3d\phi_4}{(2i\pi)^4}\D_0(\phi_1)\D_0(\phi_2)\D_0(\phi_3)\D_0(\phi_4)G_0(\phi_{12})G_0(\phi_{23})G_0(\phi_{34})G_0(\phi_{41})}=\dfrac{35q^4}{144\e_1^8}+O(q^5).
\end{split}
\end{align}
Note that a cluster of order $n$ contains at least $n$ vertices and thus provide a term of order $O(q^n)$ to the $q$-expansion. Taking the sum of these terms, we find
\begin{equation}\label{CFB}
\CF_B^{(1)}=\dfrac{7q^3}{9\e_1^6}-\dfrac{1519}{1728}\dfrac{q^4}{\e_1^8}+O(q^5).
\end{equation}
Again, the term of order $O(q^3)$ can be computed directly by evaluation of the clusters displayed in figure \refOld{3nodesB},
\begin{equation}
a)=\dfrac{3q^3}{2\e_1^6}\e_2+O(\e_2^2),\quad b)=-\dfrac{q^3}{\e_1^6}\e_2+O(\e_2^2),\quad c)=\dfrac{5q^3}{18\e_1^6}\e_2+O(\e_2^2).
\end{equation}

Summing both contributions \ref{CFA} and \ref{CFB}, the only non-vanishing term is of order $O(q)$, $\CF^{(1)}=-q/\e_1^2+O(q^5)$. Combining with $\CF^{(0)}=q/\e_1$, we recover
\begin{equation}
\CF_{U(1)}=\e_2\log\CZ_{U(1)}=\dfrac{q}{\e_1}\left(1-\dfrac{\e_2}{\e_1}+O(q^5,\e_2^2)\right)=\dfrac{\Lambda}{\e_1}+O(\Lambda^5,\e_2^2).
\end{equation}


\end{document}